\documentclass[nofootinbib,aps,11pt]{revtex4-1}
\usepackage{graphicx}
\usepackage{subfigure} 
\usepackage{hyperref}
\usepackage{cancel}
\usepackage{amssymb}
\usepackage{textcomp}
\usepackage{amsmath}
\usepackage{bm}
\usepackage{times}
\usepackage{epsfig}
\usepackage{color}
\newcommand{\hc}{{\rm h.c.}}

\newcommand{\GeV}{{\rm GeV}}

\newcommand{\SM}{{\rm SM}}
\newcommand{\eq}{{\rm eq}}
\begin{document}
\title{\LARGE Sterile Neutrino Portal Dark Matter in  $\nu$THDM }
\bigskip
\author{Ang Liu$^1$}
\author{Feng-Lan Shao$^1$}
\email{sps\_shaofl@mail.sdu.edu.cn}
\author{Zhi-Long Han$^2$}
\email{sps\_hanzl@ujn.edu.cn}
\author{Yi Jin$^2$}
\author{Honglei Li$^2$}
\affiliation{$^1$School of Physics and Physical Engineering, Qufu Normal University, Qufu, Shandong 273165, China\\
$^2$School of Physics and Technology, University of Jinan, Jinan, Shandong 250022, China}
\date{\today}
\begin{abstract}
In this paper, we propose the sterile neutrino portal dark matter in $\nu$THDM. This model can naturally generate tiny neutrino mass with the neutrinophilic scalar doublet $\Phi_{\nu}$ and sterile neutrinos $N$ around TeV scale. Charged under a $Z_2$ symmetry, one Dirac fermion singlet $\chi$ and one scalar singlet $\phi$ are further introduced in the dark sector. The sterile neutrinos $N$ are the  mediators between the DM and SM. Depending on the coupling strength, the DM can be either WIMP or FIMP. For the WIMP scenario, pair annihilation of DM into $NN$ is the key channel to satisfy various bounds, which could be tested at indirect detection experiments. For the FIMP scenario, besides the direct production of DM from freeze-in mechanism, contributions from late decay of NLSP is also important. When sterile neutrinos are heavier than the dark sector, NLSP is long-lived due to tiny mixing angle between sterile and light neutrinos. Constrains from free-streaming length, CMB, BBN and neutrino experiments are considered.
\end{abstract}
\maketitle

\section{Introduction}
For many years, the standard model of particle physics has been proved very successful. But it could not explain some phenomena, such as the origin of tiny neutrino mass and particle DM. Searching for the connections between these two parts has become an interesting topic at present, where new physics beyond the SM is required~\cite{Krauss:2002px,Asaka:2005an,Ma:2006km,Aoki:2008av,Cai:2017jrq}.

The right-hand neutrinos $N$ are introduced in the traditional Type-I seesaw mechanism \cite {Minkowski:1977sc,Mohapatra:1979ia} to solve the tiny neutrino mass problem. Although a keV-scale right-hand neutrino could both explain the active neutrino mass and act as DM candidates \cite {Dodelson:1993je,Drewes:2016upu,Datta:2021elq}, the parameter space is now tightly constrained by X-ray searches \cite{Ng:2019gch}. To avoid such constraints, one may introduce an exact $Z_2$ symmetry to make the lightest right-hand neutrino $N_1$ stable \cite{Molinaro:2014lfa,Hessler:2016kwm,Baumholzer:2018sfb,Baumholzer:2019twf}. This $Z_2$ symmetry also forbids the direct Yukawa interaction $y\overline{L}\tilde{\Phi}N$, thus neutrino masses are not appear at tree-level. Provided an additional inert scalar doublet, the right-hand neutrinos can mediate the generation of tiny neutrino mass at one-loop level, which is known as the Scotogenic model \cite{Ma:2006km}. 

Another pathway is assuming the right-hand neutrinos $N$ as the messenger between the dark sector and standard model \cite{Pospelov:2007mp,Gonzalez-Macias:2016vxy,Escudero:2016tzx,Escudero:2016ksa,Batell:2017cmf,Ballett:2019pyw,Biswas:2021kio,Coy:2021sse,Borah:2021pet,Fu:2021uoo,Coito:2022kif,Biswas:2022vkq}. With sizable coupling between DM and $N$, sufficient annihilation rate of DM into $NN$ pairs is viable. When this channel is dominant, the DM-nucleon scattering cross section is suppressed, hence satisfy the constraints from direct detection. Meanwhile, observable signature is still expected by indirect detection \cite{Tang:2015coo,Campos:2017odj,Batell:2017rol,Folgado:2018qlv}. On the other hand, the coupling between DM and $N$ can be tiny, thereby producing DM via the freeze-in mechanism \cite{Becker:2018rve,Chianese:2018dsz,Bian:2018mkl,Bandyopadhyay:2020qpn,Cheng:2020gut,Chianese:2021toe,Cheng:2021umr}.
Leptogenesis in the frame work of neutrino portal DM is also discussed in Ref.~\cite {Bernal:2017zvx,Falkowski:2017uya,Liu:2020mxj,Chang:2021ose,Barman:2021ost}.

To naturally obtain tiny neutrino mass, the right-hand neutrinos should be at the scale of $10^{14}$~GeV with $\mathcal{O}(1)$ Yukawa coupling \cite{Chianese:2019epo,Chianese:2020khl}. However, phenomenological studies usually favor the right-hand neutrinos below TeV scale. In this scenario, the Yukawa coupling $y$ is at the order of $10^{-7}$. Therefore, the right-hand neutrinos may not be fully thermalized \cite{Tang:2016sib,Bandyopadhyay:2018qcv}, which affects the relic density of DM. TeV scale right-hand neutrinos with large Yukawa coupling $y$ is possible in low-scale seesaw, e.g., inverse seesaw \cite{Mohapatra:1986aw,Mohapatra:1986bd} and neutrinophilic two Higgs doublet model ($\nu$THDM) \cite{Ma:2000cc}.

In this paper, we propose the sterile neutrino portal DM in the $\nu$THDM. A neutrinophilic Higgs doublet $\Phi_\nu$ with lepton number $L_{\Phi_\nu}=-1$ is introduced. By assuming the right-hand neutrinos with vanishing lepton number $L_N=0$, they couple exclusively to $\Phi_{\nu}$ via the Yukawa interaction $\overline{L}\tilde{\Phi}_\nu N$, while interaction with standard model Higgs doublet $\Phi$ is forbidden. The soft term $\mu^2 \Phi^\dag\Phi_{\nu}$ then induces a naturally small vacuum expectation value of $\Phi_{\nu}$, resulting light right-hand neutrinos with large Yukawa coupling. The dark sector consists of one Dirac fermion singlet $\chi$ and one scalar singlet $\phi$. Both are charged under a $Z_2$ symmetry, of which the lightest one serves as DM candidate. As a mediator, the right-hand neutrinos couple with the dark sector through the Yukawa interaction $\lambda_{ds} \bar{\chi}\phi N$. Depending on the strength of $\lambda_{ds}$ and other relevant couplings, both WIMP and FIMP DM scenario are possible. A comprehensive analysis of the DM phenomenology is carried out in the following studies.

The structure of the paper is organized as follows. In Sec. \ref{SEC:TM}, we briefly introduce the model. A detail study of the WIMP scenario in aspect of relic density, Higgs invisible decay, direct and indirect detection is performed in Sec. \ref{SEC:WIMP}. Then we study the FIMP scenario in Sec. \ref{SEC:FIMP}. Finally, we summarize our results in Sec. \ref{SEC:CL}.

\section{The Model}\label{SEC:TM}

\begin{table}
	\begin{center}\large
		\begin{tabular}{|c| c c c | c c c |} 
			\hline
		 			&  $L$  & ~$N$~ & ~$\chi$~ & ~$\Phi$~ & ~$\Phi_\nu$~ & ~$\phi$~ \\ \hline
		 $SU(2)_L$ & 2 & 1 & 1 & 2 & 2 & 1\\ \hline
		 $U(1)_Y$  & $-\frac{1}{2}$~  & 0 & 0 & $\frac{1}{2}$ & $\frac{1}{2}$ & 0 \\ \hline
		 $U(1)_L$  & 1 & 0 & 0 & 0 & $-1$~ & 0\\\hline
		 $Z_2$     & $+$ & $+$ & $-$ & $+$ & $+$ & $-$ \\
			\hline
		\end{tabular}
	\end{center}
	\caption{Relevant particle contents and corresponding charge assignments. 
		\label{Tab:Particle}}
\end{table}

This model extends the $\nu$THDM with a dark sector. The particle contents and corresponding charge assignments are listed in table \ref{Tab:Particle}. Under the global $U(1)_L$ symmetry, right-hand neutrinos can not couple to standard Higgs doublet $\Phi$, but to the new one $\Phi_{\nu}$. The $Z_2$ symmetry is exact to stabilize DM. In this paper, both the fermion and scalar DM are considered. The scalar potential under above symmetry is given by
 \begin{eqnarray} \label{sp}
 	V & = & m_{\Phi}^2 \Phi^\dag \Phi +  m_{\Phi_\nu}^2 \Phi^\dag_\nu \Phi_\nu
 	+m_\phi^{\prime 2} \phi^\dag\phi+\frac{\lambda_1}{2} (\Phi^\dag \Phi)^2+\frac{\lambda_2}{2} (\Phi^\dag_\nu \Phi_\nu)^2 \nonumber \\
 	& & +\lambda_3 (\Phi^\dag \Phi)(\Phi^\dag_\nu \Phi_\nu)+\lambda_4(\Phi^\dag \Phi_\nu)(\Phi^\dag_\nu \Phi)   - (\mu^2 \Phi^\dag\Phi_\nu +\hc ) \nonumber \\ 
 	&& +\frac{\lambda_5}{2} (\phi^\dag\phi)^2 + \lambda_6 (\phi^\dag \phi)(\Phi^\dag\Phi)+\lambda_7 (\phi^\dag \phi) (\Phi_\nu^\dag\Phi_\nu),
 \end{eqnarray}
where the $\mu^2$ term breaks the $U(1)_L$ symmetry explicitly but softly. Since this term is the only source of $U(1)_L$ breaking, it is naturally small and stable under radiative corrections \cite{Haba:2011fn,Davidson:2009ha}. The small $\mu^2$ term may originate from the spontaneous breaking of global $U(1)_L$ \cite{Wang:2016vfj}. The electroweak symmetry is broken spontaneously by $\Phi$ as usual. With the condition $m_{\Phi}^2<0,  m_{\Phi_\nu}^2>0,  |\mu^2|\ll m_{\Phi_\nu}^2$, one could be obtained hierarchical vacuum expectation values as
\begin{equation}
	v\simeq \sqrt{\frac{-2 m_{\Phi}^2}{\lambda_1}},
	 v_\nu \simeq \frac{\mu^2 v}{m_{\Phi_\nu}^2+(\lambda_3+\lambda_4)v^2/2},
\end{equation}
where $v=246$ GeV, and $v_\nu\sim10$ MeV with $\mu\sim1$ GeV, $m_{\Phi_{\nu}}\sim200$ GeV typically. Notably, the constraints from lepton flavor violation require $v_\nu\gtrsim 1$ MeV \cite{Guo:2017ybk}.

After the symmetry breaking, we have six physical Higgs bosons. They are the standard model Higgs boson $h$ \cite{Aad:2012tfa,Chatrchyan:2012xdj}, the charged Higgs $H^\pm$, the CP-even Higgs $H$, the CP-odd Higgs $A$, and the inert Higgs $\phi$.
Ignoring the terms of $\mathcal{O}(v_\nu^2)$ and $\mathcal{O}(\mu^2)$, their masses are
 \begin{eqnarray}
 	m_h^2 \simeq  \lambda_1 v^2,~
 	m_{H^\pm}^2\simeq m_{\Phi_\nu}^2\!+\frac{1}{2}\lambda_3v^2,~m_A^2\simeq m_H^2\simeq m_{H^\pm}^2\!+\frac{1}{2}\lambda_4v^2,~m_\phi^2\simeq m_\phi^{\prime 2}+\frac{1}{2}\lambda_6 v^2.
 \end{eqnarray}
For simplicity, we adopt a degenerate mass spectrum of $\Phi_ \nu$, $m_{H^+}\!=\!m_{H}\!=\!m_A\!=\!m_{\Phi_\nu}$, which is allowed by various constraints \cite{Machado:2015sha}. Fixing $v_\nu=10$ MeV, then $H^\pm\to tb/cb$ is the dominant decay channel when $m_N>m_{\Phi_\nu}$, while $H^\pm\to \ell^\pm N$ becomes the dominant one when $m_N<m_{\Phi_\nu}$ \cite{Haba:2011nb}. Anyway, $m_{\Phi_\nu}=200$~GeV with $v_\nu=10$ MeV is still allowed by current collider search \cite{ATLAS:2021upq}.

Besides providing masses for light neutrinos via Yukawa coupling with $\Phi_\nu$ and lepton doublet $L$, the right-hand neutrinos also contribute to the production of DM through coupling with dark particles $\phi$ and $\chi$.  The the new Yukawa interaction and  mass terms can be written as
\begin{equation}\label{yuk}
	-\mathcal{L}_Y\supset y\overline{L} \widetilde{\Phi}_\nu N  +\lambda_{ds} \bar{\chi}\phi N  +\frac{1}{2}\overline{N ^c}m_{N } N  + m_\chi \bar{\chi} \chi+ \hc,
\end{equation}
where $\widetilde{\Phi}_\nu=i\sigma_2 \Phi_\nu^*$. The tiny neutrino mass is generated via Type-I seesaw like mechanism with $\Phi$ simply replaced by $\Phi_\nu$,
\begin{equation}\label{eq:mv}
	m_\nu = - \frac{v_\nu^2}{2} y~ m_{N}^{-1} y^T.
\end{equation}
$m_\nu\sim0.1$ eV can be obtained with $v_\nu\sim10$ MeV, $y\sim0.01$ and $m_N\sim200$ GeV. With such large Yukawa coupling, the right-hand neutrinos are fully thermalized through interaction with $\Phi_{\nu}$ and $L$. Therefore, the conventional calculation of relic density are not affected in our studies.

In this paper, we focus on the DM phenomenology.  For sufficient large Yukawa coupling $\lambda_{ds}$ or other relevant couplings, the DM is a WIMP candidate produced via the freeze-out mechanism. In contrast,  tiny $\lambda_{ds}$ and other relevant couplings lead to FIMP candidate produced via the freeze-in mechanism. These two scenarios are both considered in the following studies. The complete model file is written by FeynRules package \cite {Alloul:2013bka}, then micrOMEGAs \cite{Belanger:2013oya}  is used to calculate the relic density, DM-nucleon scattering and annihilation cross section.

\section{WIMP Dark Matter} \label{SEC:WIMP}

\subsection{Relic Density}

\begin{figure}
	\begin{center}
		\includegraphics[width=0.9\linewidth]{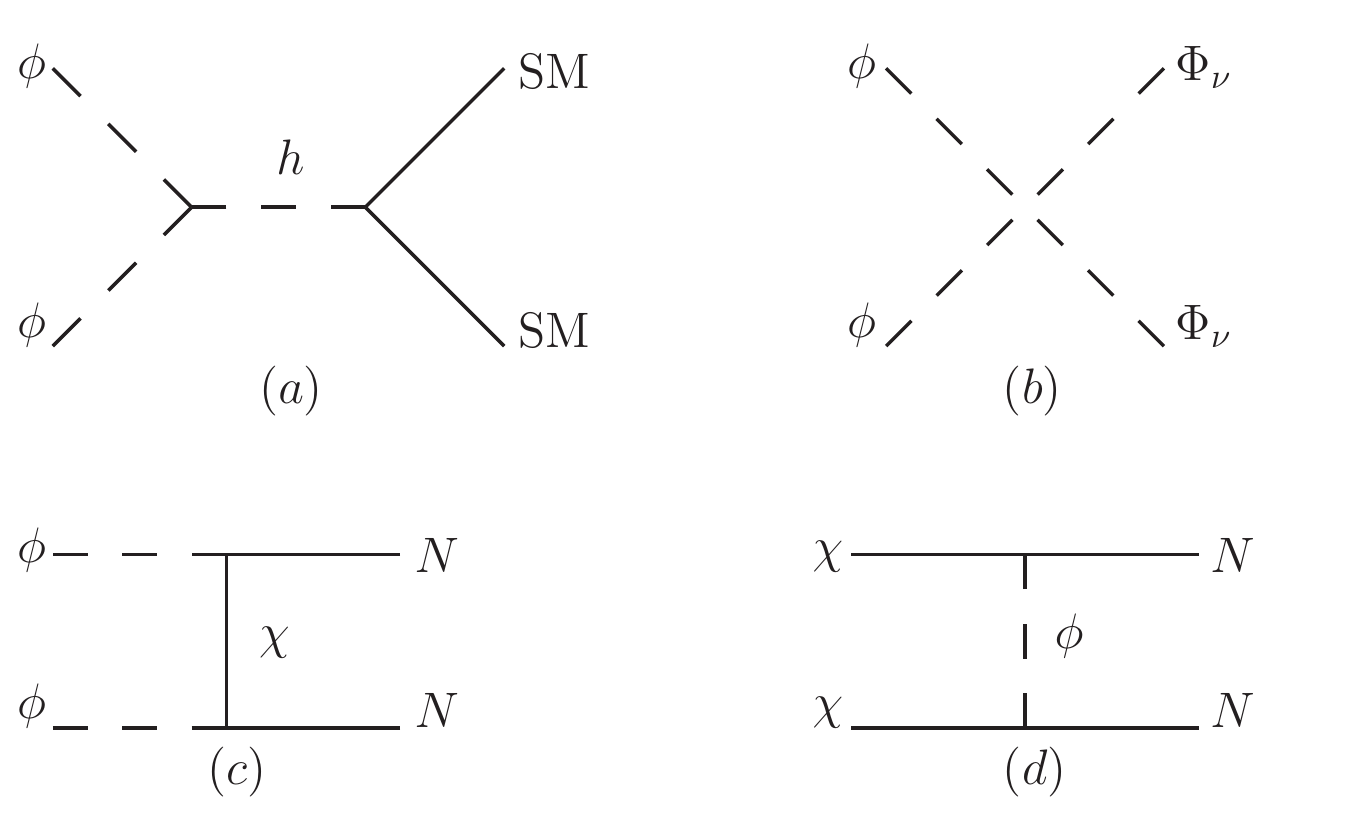}	
	\end{center}
	\caption{The annihilation channels for WIMP DM. Feynman diagrams (a)-(c) are for scalar DM, and Feynman diagram (d) is for fermion DM.}
	\label{FIG:WFeyn}
\end{figure}

When $m_\phi<m_\chi$, the scalar $\phi$ is treated as the DM candidate. In the case of WIMP DM, the most common way for scalar DM $\phi$ is annihilating to the SM final state via the standard Higgs portal coupling term $\lambda_6 (\phi^\dag \phi)(\Phi^\dag\Phi)$, which has been extensively studied in previous paper \cite{McDonald:1993ex,Burgess:2000yq,Cline:2013gha,Feng:2014vea,Arcadi:2019lka}. In our scenario, there are two extra ways for scalar DM annihilation. One is annihilating to neutrinophilic scalars $\Phi_{\nu}^\dag\Phi_{\nu}$ through the coupling $\lambda_7 (\phi^\dag \phi) (\Phi_\nu^\dag\Phi_\nu)$. The other one is annihilating to sterile neutrinos $NN$ via the Yukawa interaction $\lambda_{ds} \bar{\chi}\phi N$. The fermion DM is realized with $m_\chi<m_\phi$. It can only annihilate to sterile neutrinos $NN$ with a $t$-channel mediator $\phi$. Relevant annihilation channels are shown in Fig.~\ref{FIG:WFeyn}.

According to the nature of WIMP, DM is in thermal equilibrium with other particles in the early universe. Then it starts to freeze-out and decouple from thermal bath as the temperature decreases. The evolution of abundance $Y$ is obtained by solving the Boltzmann equation
\begin{eqnarray}\label{y-w}
	\frac{dY}{dz} = - \frac{k}{z^2}\left \langle \sigma v \right \rangle (Y^2-Y_{\text{eq}}^{2}),
\end{eqnarray}
where we use the definition $z\equiv m_\text{DM}/T$. Here, $m_\text{DM}$ is the mass of DM and $T$ is the temperature. The parameter $k$ is denoted as $k=\sqrt{\pi g_\star/45}m_\text{DM} M_{Pl}$, where $g_{\star}$ is the effective number of degrees of freedom of the relativistic species and $M_{Pl}=1.2 \times 10^{19}$ GeV is the Planck mass. 

\begin{figure}
	\begin{center}
		\includegraphics[width=0.45\linewidth]{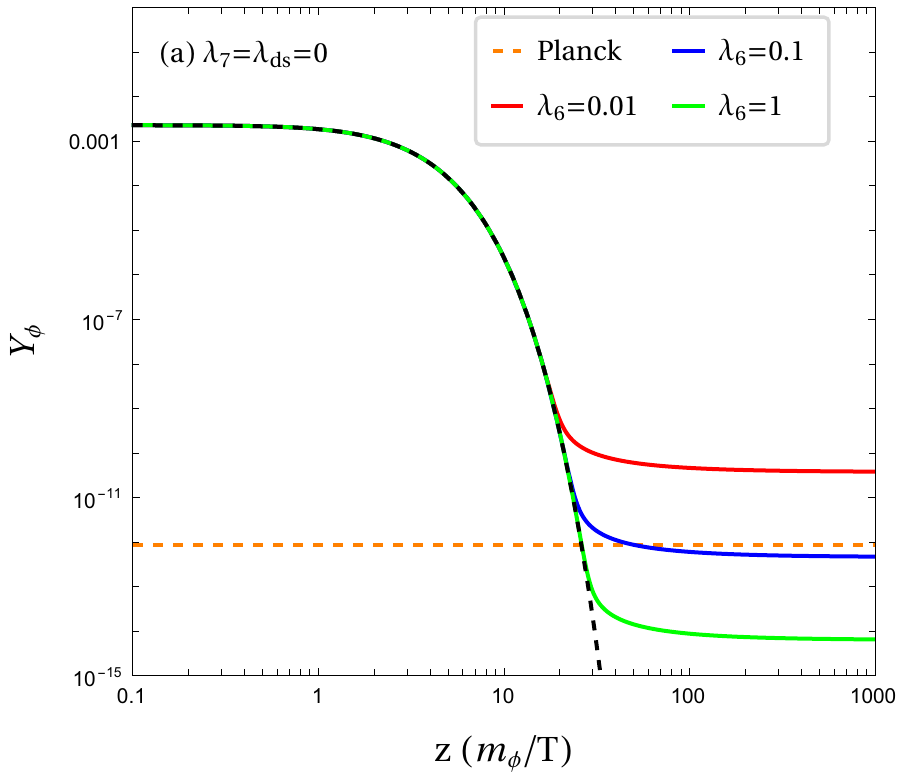}
		\includegraphics[width=0.45\linewidth]{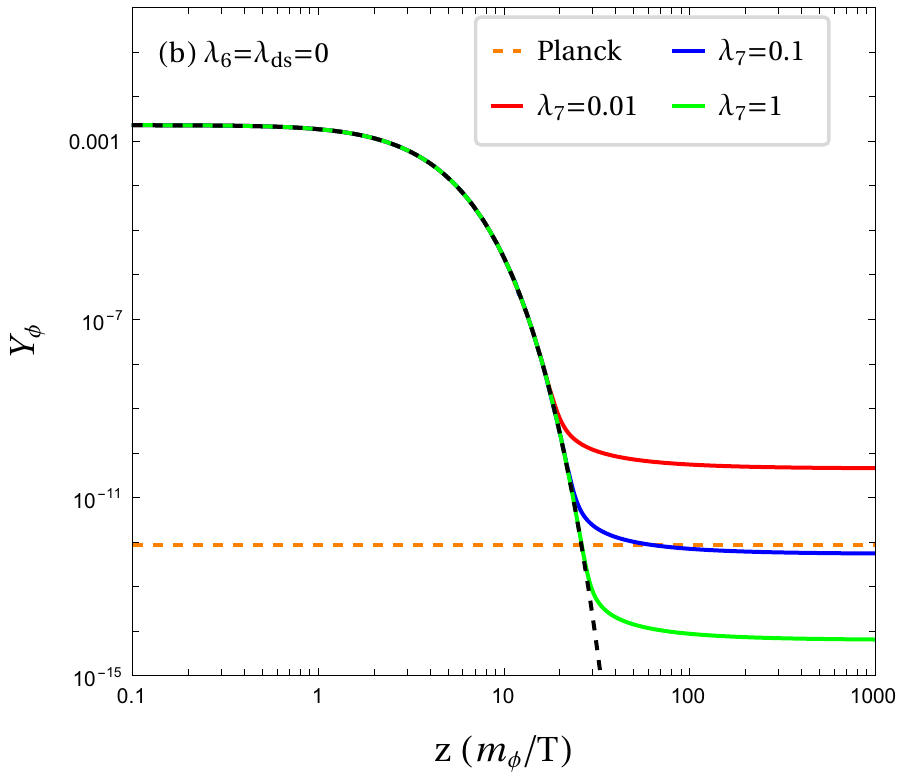}
		\includegraphics[width=0.45\linewidth]{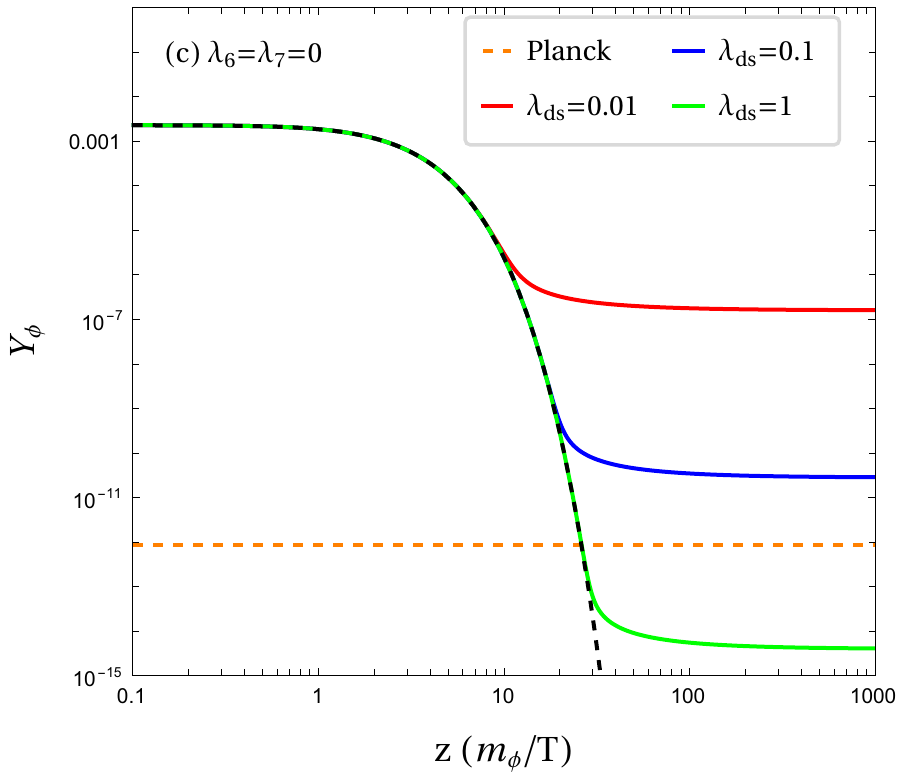}
		\includegraphics[width=0.45\linewidth]{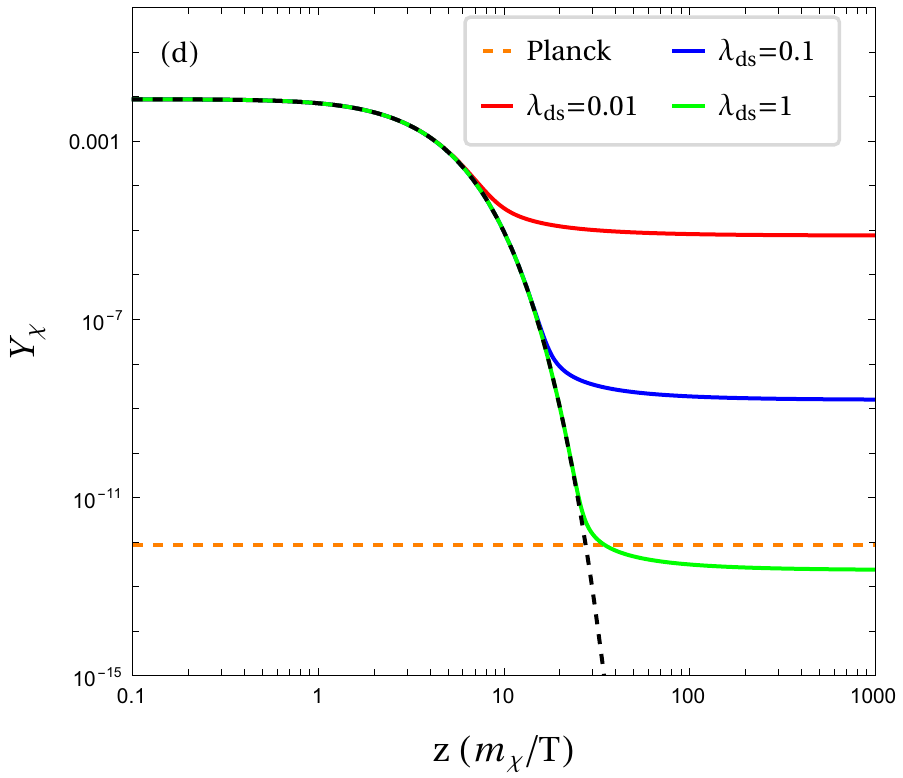}
	\end{center}
	\caption{ The evolution of DM abundance for the four annihilation channels. Subfigures (a)-(c) are for scalar DM, and subfigure (d) is for fermion DM. The orange horizontal lines predict the  Planck observed relic density $\Omega_\text{DM}h^2=0.12$ \cite{Planck:2018vyg} for $m_\text{DM}=500$ GeV. The black dashed lines describe the evolution of $Y_\text{eq}$. }
	\label{FIG:WAbun}
\end{figure}

The evolution of DM abundance for various benchmark scenarios are shown in Fig.~\ref{FIG:WAbun}, which correspond to the individual contribution of annihilation channels in Fig.~\ref{FIG:WFeyn}. During the calculation, we fix $m_N=250$ GeV, $m_\text{DM}=500$ GeV, and $|m_\phi-m_\chi|=500$ GeV. When the scalar DM $\phi$ annihilates via the Higgs or neutrinophilic scalar portal channels, correct relic density is predicted with couplings $\lambda_{6,7}\sim0.1$. When the sterile neutrino portal becomes the dominant channel, the coupling $\lambda_{ds}$ should be slightly larger than 0.1 to realize correct relic density. Meanwhile, $\lambda_{ds}\sim1$ is required for fermion DM $\chi$ with correct relic density. 

\begin{figure}
	\begin{center}
		\includegraphics[width=0.45\linewidth]{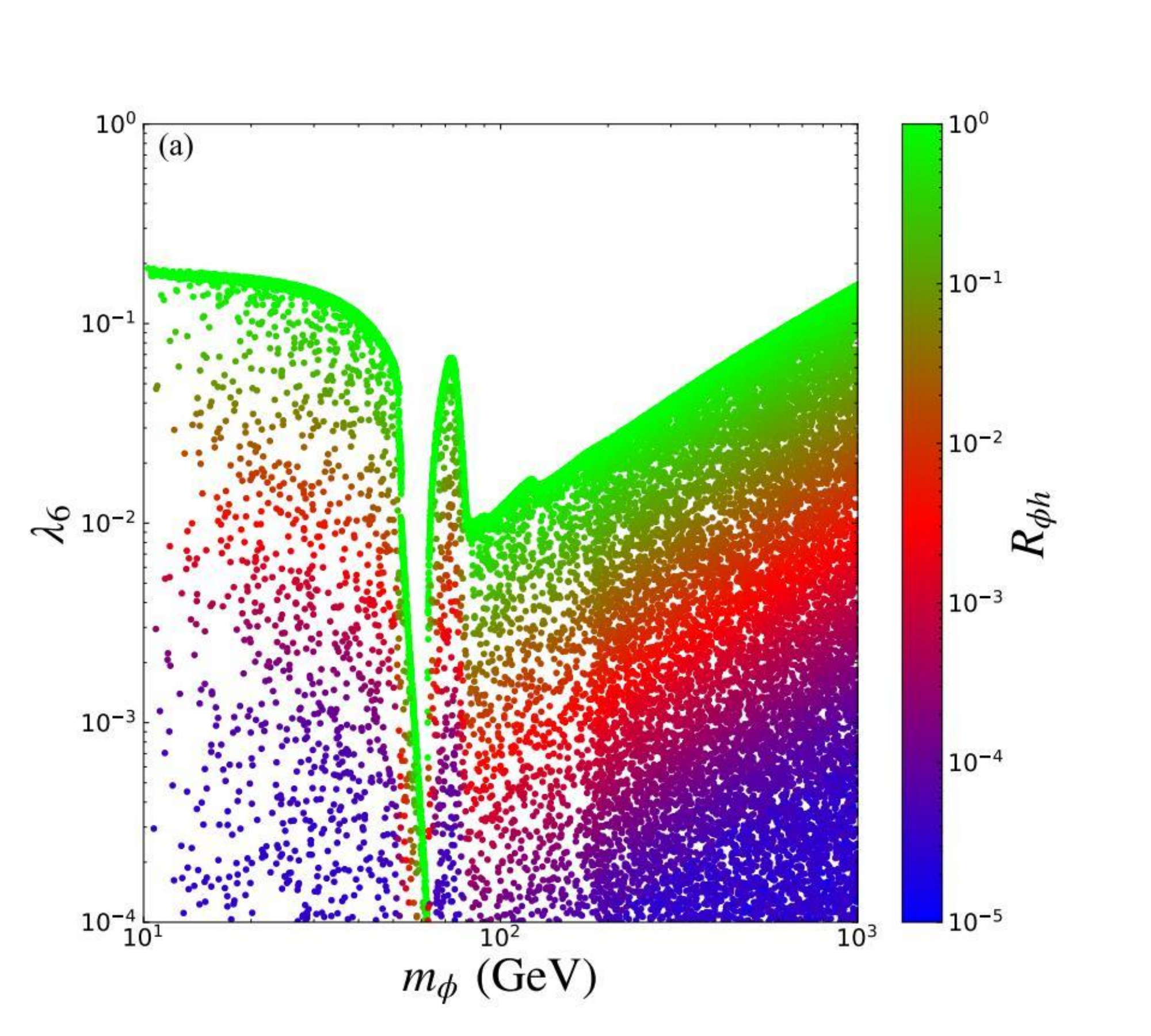}
		\includegraphics[width=0.45\linewidth]{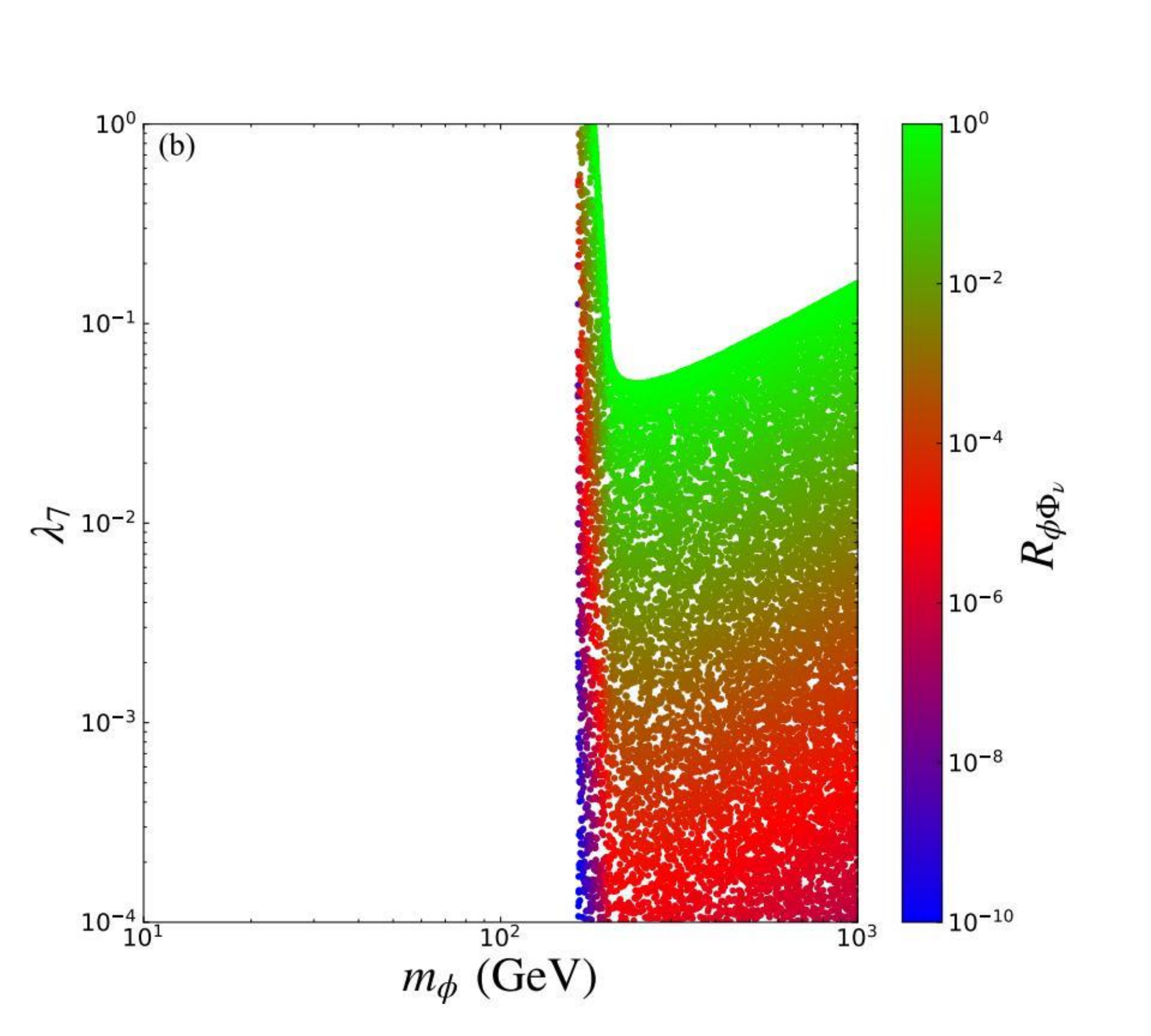}
		\includegraphics[width=0.45\linewidth]{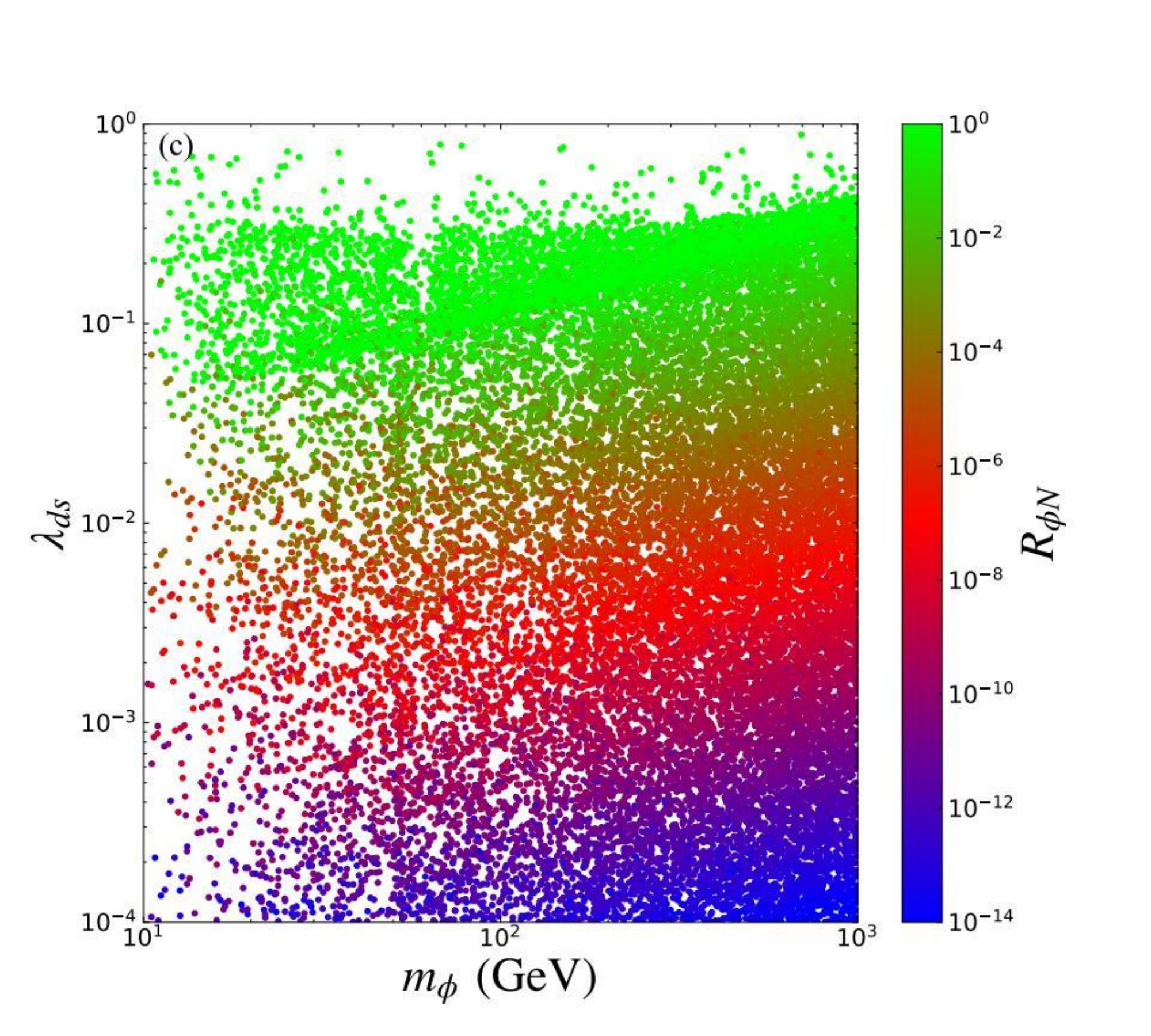}
		\includegraphics[width=0.45\linewidth]{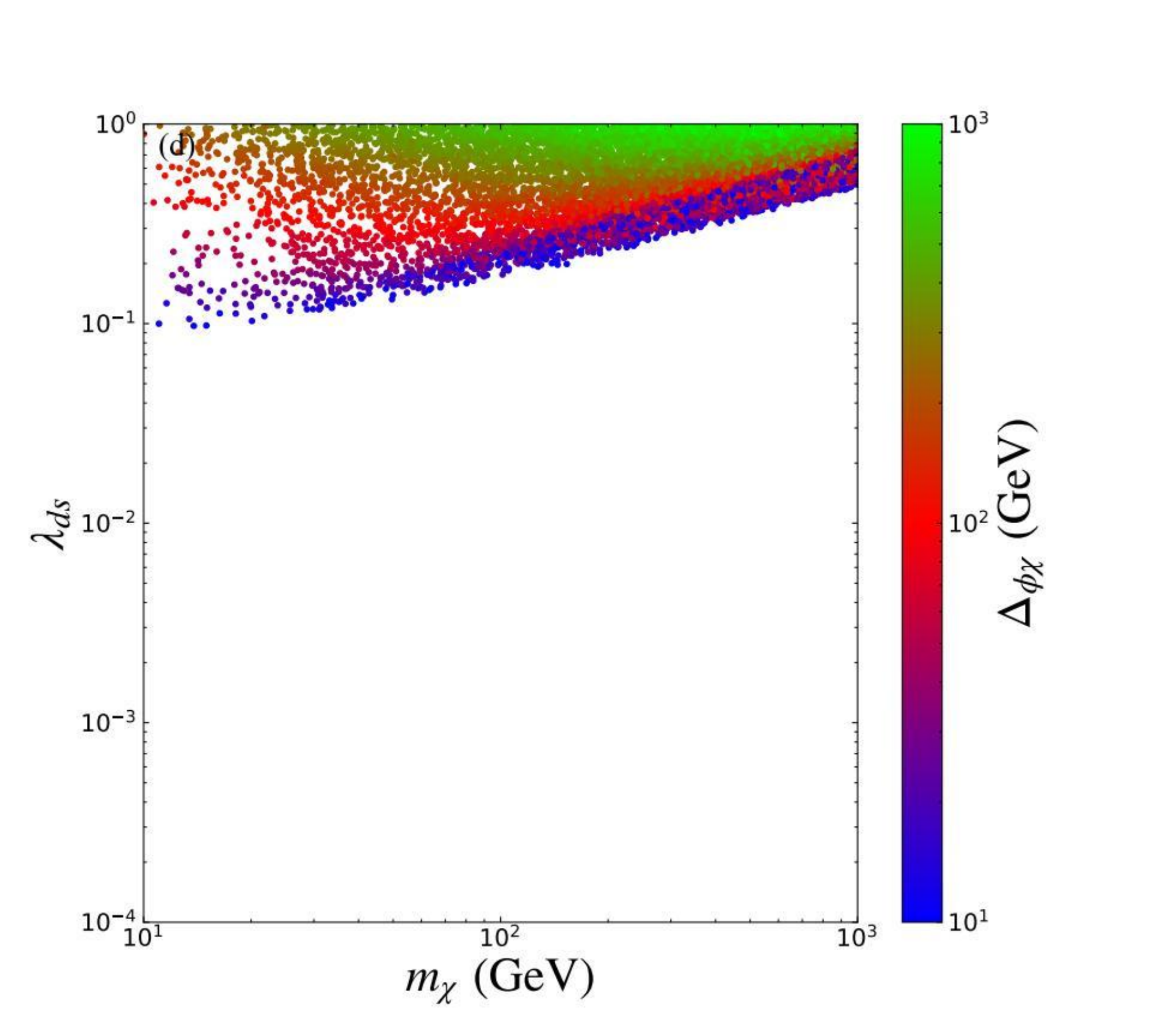}
	\end{center}
	\caption{ Allowed parameter space with correct relic density. Relative contributions of various annihilation channels to the total abundance of the scalar DM $\phi$ are illustrated in subfigure (a)-(c). Subfigure (d) is for fermion DM $\chi$, which is colored by the mass splitting $\Delta_{\phi\chi}=m_\phi-m_\chi$.}
	\label{FIG:WRelic}
\end{figure}

To obtain the parameter space with correct relic density in $3\sigma$ range of the Planck observed result \cite{Planck:2018vyg}, i.e., $\Omega_\text{DM} h^2\in[0.117,0.123]$, we perform a random scan on relevant parameters in the following range based on previous benchmark scenarios
\begin{eqnarray}\label{s-w-scope}
	\begin{aligned}
		m_{\phi,\chi,N}\in[10,1000]~\GeV, \lambda_{6,7,ds}\in[0.0001,1].
	\end{aligned}
\end{eqnarray}
Meanwhile, we fix the masses of neutrinophilic scalars as $m_{\Phi_{\nu}}=200$ GeV for simplicity. The results are shown in Fig.~\ref{FIG:WRelic}, where the relative contribution of annihilation channels of the scalar DM are defined as
\begin{eqnarray}
	R_{\phi h}&=&\frac{\Omega_{\rm DM}h^2}{\Omega_{\phi \SM}h^2},	R_{\phi \Phi_\nu}=\frac{\Omega_{\rm DM}h^2}{\Omega_{\phi \Phi_{\nu}}h^2}, 	R_{\phi N}=\frac{\Omega_{\rm DM}h^2}{\Omega_{\phi N}h^2}.
\end{eqnarray}
Here, $\Omega_{\phi \SM}h^2$, $\Omega_{\phi \Phi_{\nu}}h^2$ and $\Omega_{\phi N}h^2$ denote the relic density for the SM Higgs portal $\phi\phi\to \SM \overline{\SM}$, neutrinophilic scalar portal $\phi\phi\to \Phi_\nu \Phi_{\nu}^\dag$ and sterile neutrino portal $\phi\phi\to NN$ channel respectively. As for the fermion DM, since the sterile neutrino portal $\chi\bar{\chi}\to NN$ is the only possible annihilation channel, we instead color the samples by the mass splitting $\Delta_{\phi\chi}=m_\phi-m_\chi$.

For the scalar DM, green samples indicate that the corresponding annihilation channel has a relative large contribution to the total abundance. In contrast, contributions of the red and blue samples are negligible. It is clear in Fig.~\ref{FIG:WRelic} (a) that there is a sharp dip around $m_\phi\sim m_h/2$ in the $\lambda_{6}-m_\phi$ plane due to the resonance production of SM Higgs $h$ in the $s$-channel. Typically, the $\phi\phi\to \Phi_\nu \Phi_{\nu}^\dag$ and the $\phi\phi\to NN$ channel become the dominant one with $\lambda_7\gtrsim0.01$ and $\lambda_{ds}\gtrsim0.1$ respectively. When $m_\phi$ is close to $m_{\Phi_\nu}$, the cross section is suppressed by the final state phase space, thus a relative larger coupling $\lambda_7$ is required. As for the fermion DM, $\lambda_{ds}\gtrsim0.1$ should be satisfied to generate correct abundance. The lower bound on $\lambda_{ds}$ becomes larger when $m_\chi$ is heavier. It is also obvious that a smaller mass splitting $\Delta_{\phi \chi}$ is accompanied by a smaller $\lambda_{ds}$.

\subsection{Higgs Invisible Decay}

Now, we consider the possible constraints on the above parameter space for correct abundance. When the SM Higgs decays into DM, the invisible Higgs decay rate will be enhanced. Currently, the invisible branching ratio is constrained  by the ATLAS experiment \cite{ATLAS:2020kdi}, namely, ${\rm BR}_{\rm inv}=\Gamma_{\rm inv}/\left(\Gamma_{\rm inv}+\Gamma_{\rm SM}\right)\textless 0.11$, where the SM Higgs width is $\Gamma_{\rm SM} \approx 4 $ MeV. In the case of scalar DM $\phi$, the invisible Higgs decay  $h\rightarrow\phi\phi$ via the coupling $\lambda_6 (\phi^\dag \phi)(\Phi^\dag\Phi)$ is kinematically allowed when $m_\phi \textless m_h/2$.  The resulting invisible decay width can be expressed as
\begin{eqnarray}\label{hss} 
	\Gamma_{h\rightarrow {\phi\phi}}=\frac {\lambda_6^{2}v^{2}}{8\pi{m_h}}\sqrt{1-\frac{4m_\phi^2}{m_h^2}}.
\end{eqnarray}

In the case of fermion DM $\chi$, although $\chi$ does not couples to $h$ at tree level, the dimension-5 operator $(\bar{\chi}\chi)({\Phi^\dagger}\Phi)$ can generate an effective coupling between $\chi$ and $h$ at one-loop level. The partial decay width of $h\rightarrow{\chi\bar{\chi}}$ with the condition of $m_\chi<m_h/2$ can be written as
\begin{eqnarray} 
	\Gamma_{h\rightarrow {\chi\bar{\chi}}}=\frac {\lambda_{h\chi}^{2}{m_h}}{8\pi}{\left(1-\frac{4m_\chi^2}{m_h^2}\right)^{3/2}},
\end{eqnarray}
with the effective coupling $\lambda_{h\chi}$ expressed as \cite{Gonzalez-Macias:2016vxy} 
\begin{eqnarray} \label{h-ff}
	\lambda_{h\chi}=\frac {{\lambda_6}\lambda_{ds}^{2}}{16{\pi}^2}\left(\frac{m_N}{m_\phi^2-m_N^2}+\frac{2m_N^3\log\frac{m_N}{m_\phi}}{(m_\phi^2-m_N^2)^2}\right).
\end{eqnarray}

\begin{figure}
	\begin{center}
		\includegraphics[width=0.45\linewidth]{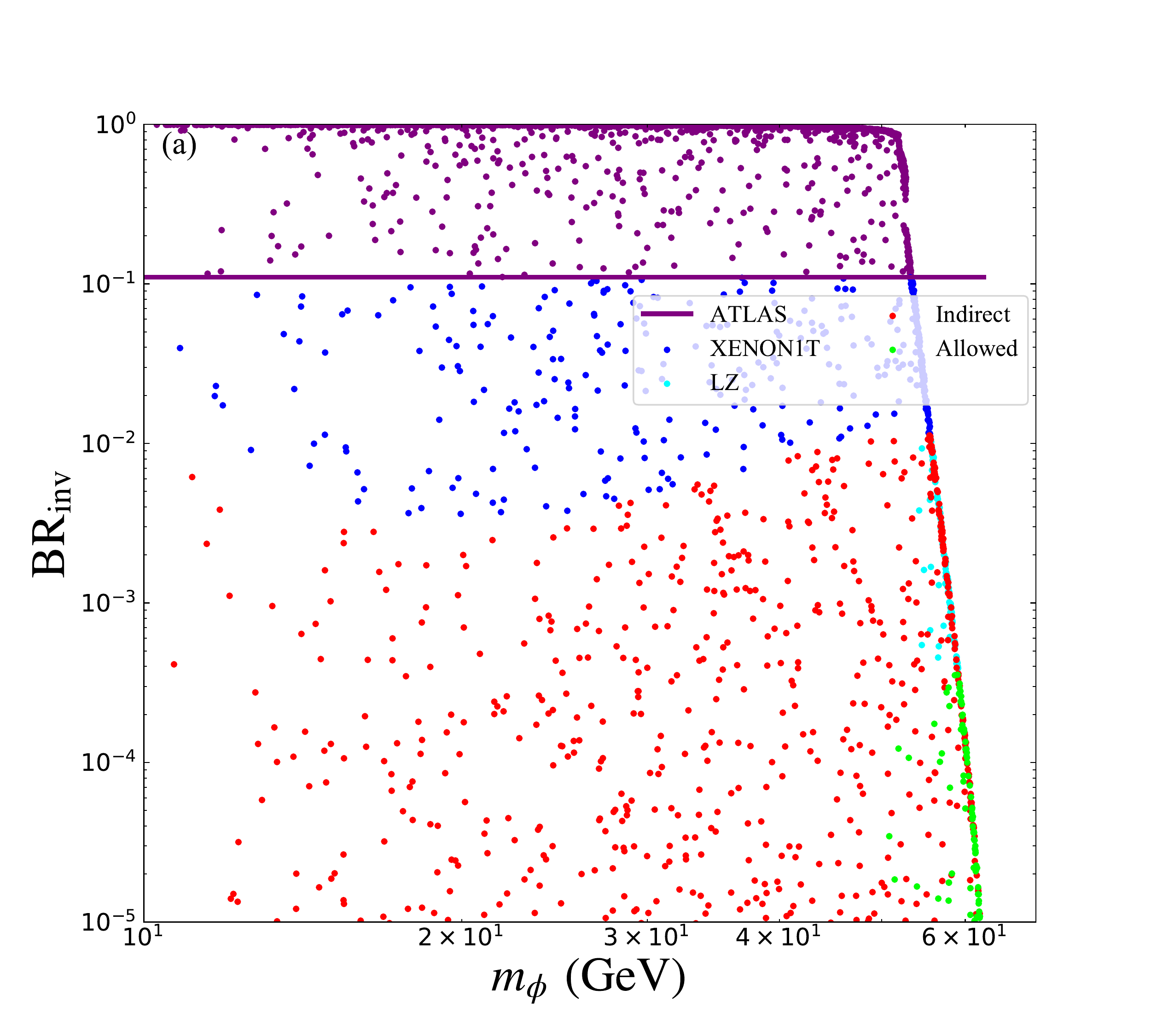}
		\includegraphics[width=0.45\linewidth]{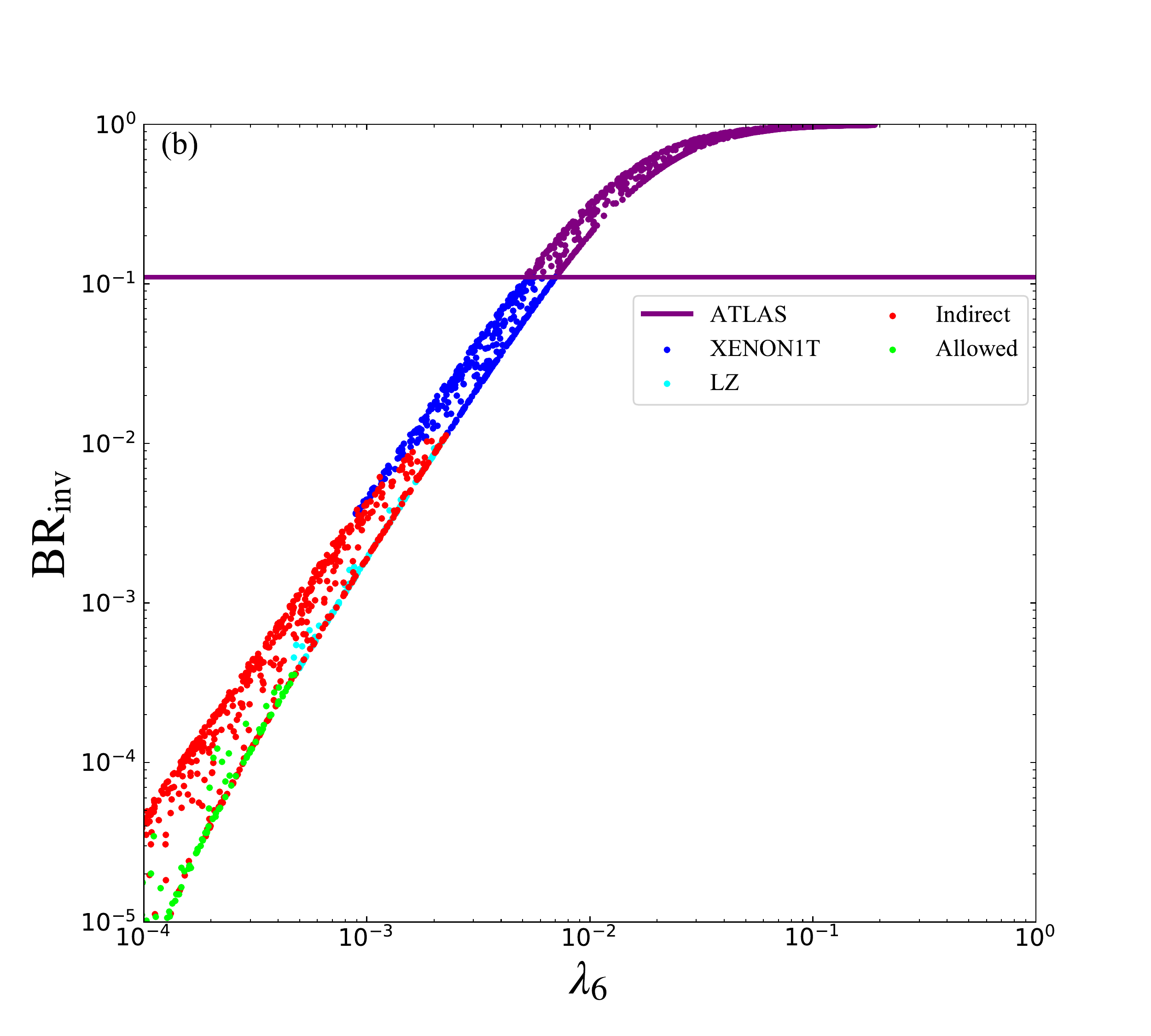}
		\includegraphics[width=0.45\linewidth]{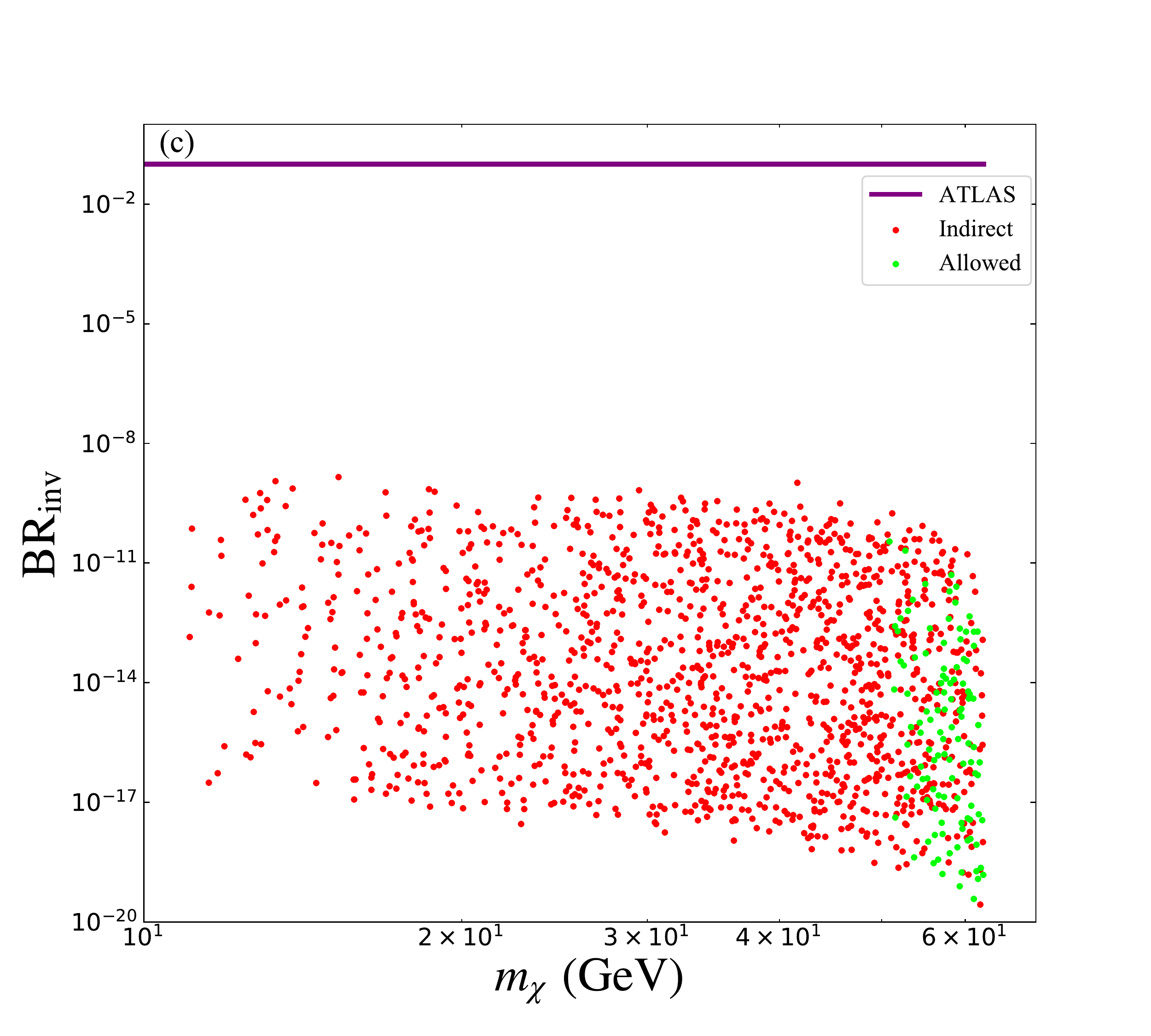}
		\includegraphics[width=0.45\linewidth]{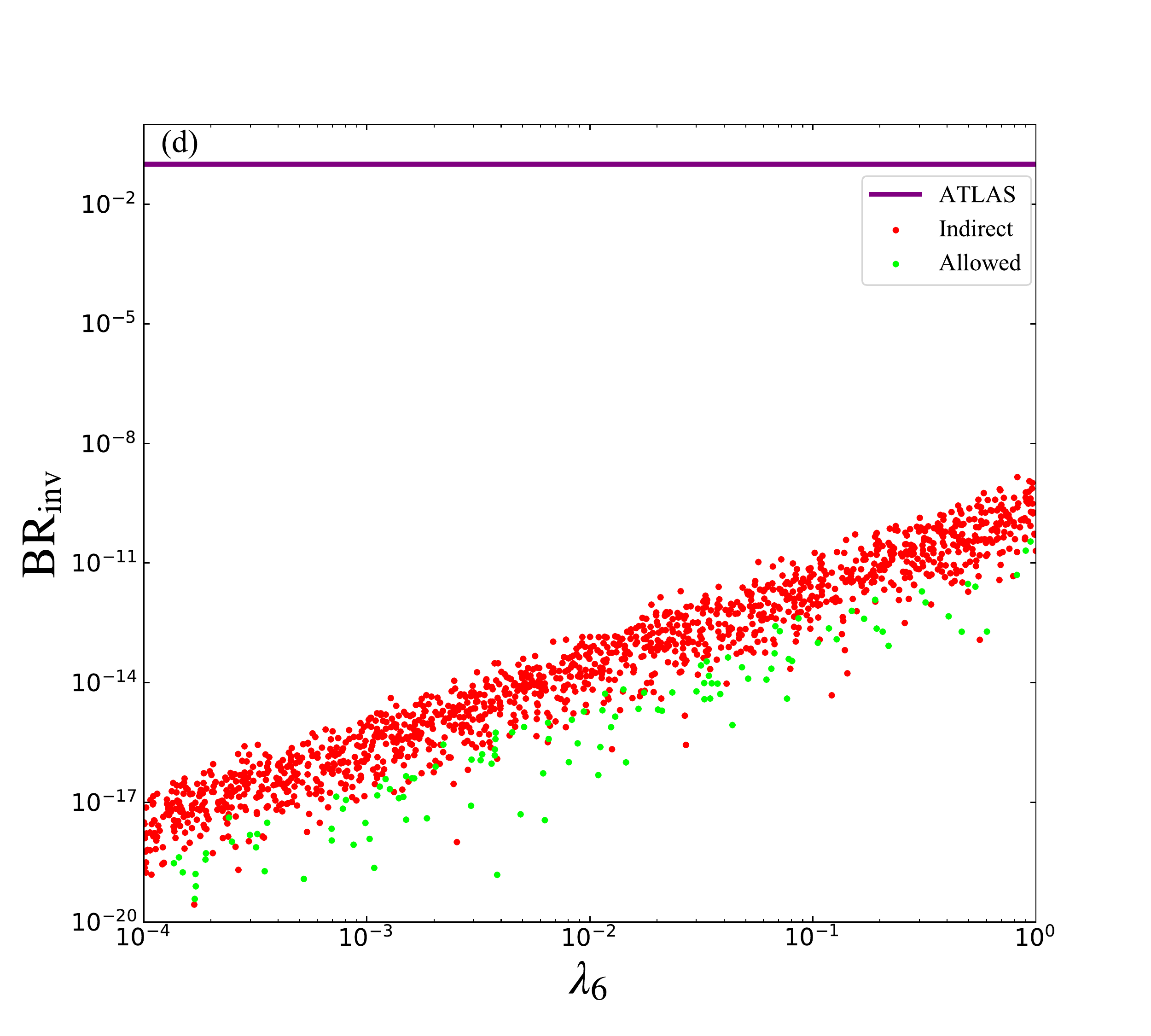}
	\end{center}
	\caption{Constraints from invisible Higgs decay. The upper(lower) two panels are for scalar(fermion) DM. The purple, blue, cyan and red points express the parameter spaces excluded by ATLAS Higgs invisible result, direct detection experiment XENON1T \cite{XENON:2018voc}, future LZ experiment\cite{LZ:2015kxe, McKinsey:2016xhn} and indirect detection experiments \cite{Batell:2017rol}, respectively. The green points satisfy all constraints mentioned. For fermion DM, the direct detection experiments XENON1T or even LZ do not exclude any points for the parameter space scanned in Eq.~\ref{s-w-scope}.}
	\label{FIG:HigInv}
\end{figure}

The Higgs invisible decay constraint on the parameter spaces for correct abundance are shown in Fig.~\ref{FIG:HigInv}.  The disallowed parameter space of scalar DM $\phi$ satisfy $\lambda_6\gtrsim 0.01$ with the constraint of ATLAS result. However, current constraint from direct detection experiment XENON1T is approximately one order of magnitude tighter than Higgs invisible decay. Moreover, the indirect detection experiments have already exclude the region $m_\phi\lesssim60$ GeV in this sterile neutrino portal model. The current allowed samples predict $\rm BR_{inv}\lesssim10^{-2}$ in the $h\to \phi\phi$ decay mode. If no signal is observed in the future LZ experiments,  $\rm BR_{inv}\lesssim3\times10^{-4}$ in this mode is expected, which will be smaller than the SM value $\rm BR_{inv}\sim10^{-3}$ in the $h\to ZZ^*$ with $Z\to \nu\bar{\nu}$ mode. For fermion DM $\chi$, the Higgs invisible decay mode $h\to \chi\bar{\chi}$ is heavily suppressed by the loop-factor. In the scanned parameter space, it has a range of $\rm BR_{inv}\lesssim 10^{-9}$, which is at least eight orders of magnitude smaller than current measured bound. Hence, no sample is excluded by Higgs invisible decay for fermion DM.

\subsection{Direct Detection}

Nowadays, the DM direct detection experiments have already set tight constraints on the  DM-nucleon scattering cross sections. Results from the XENON1T \cite{XENON:2018voc} and the future LZ \cite{LZ:2015kxe, McKinsey:2016xhn} experiments are considered in this paper. The spin-independent scattering cross section of scalar DM $\phi$ is mediated by SM Higgs $h$, which originates from the Higgs portal coupling  $\lambda_6(\phi^\dag \phi)(\Phi^\dag \Phi)$. The result of fermion DM $\chi$ is also mediated by $h$ with the one-loop generated effective coupling $\lambda_{h\chi} (\bar{\chi}\chi)(\Phi^\dag \Phi)$, where $\lambda_{h\chi}$ is given in Eq.~\ref{h-ff}. The corresponding scattering cross sections are then calculated as \cite{Arcadi:2019lka}
\begin{eqnarray}
	\sigma_\text{SI}^{\phi}&=&\frac{\lambda_6^2 f_n^2}{\pi m_h^4} \frac{m_n^4}{(m_\phi+m_n)^2}, \\
	\sigma_\text{SI}^{\chi}&=&\frac{\lambda_{h\chi}^2 f_n^2}{\pi m_h^4} \frac{m_n^4 m_\chi^2}{(m_\phi+m_n)^2},
\end{eqnarray}
where $f_n\approx0.3$ is the nucleon matrix element and $m_n\simeq0.939$ GeV is the averaged nucleon mass.

\begin{figure}
	\begin{center}
		\includegraphics[width=0.45\linewidth]{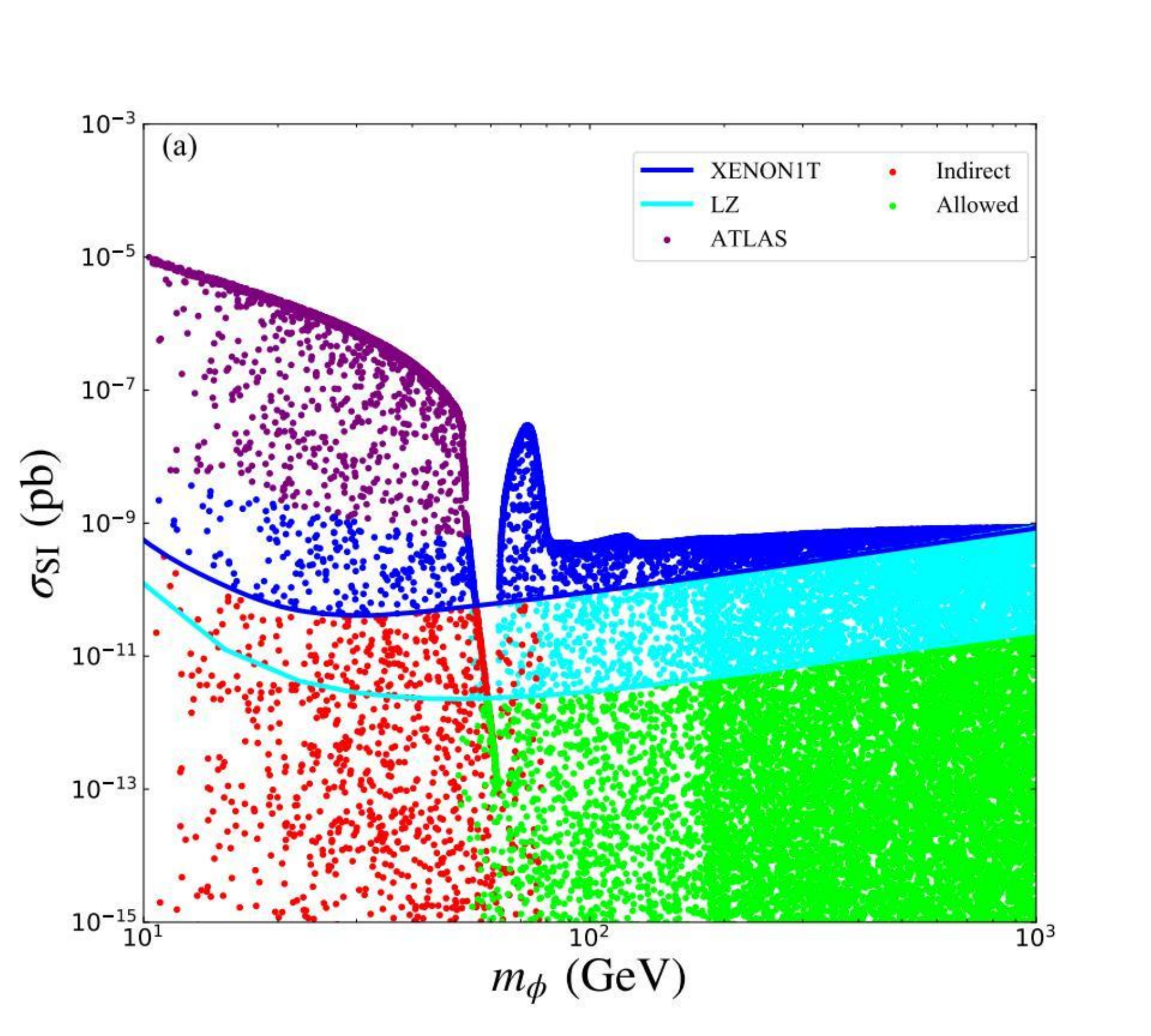}
		\includegraphics[width=0.45\linewidth]{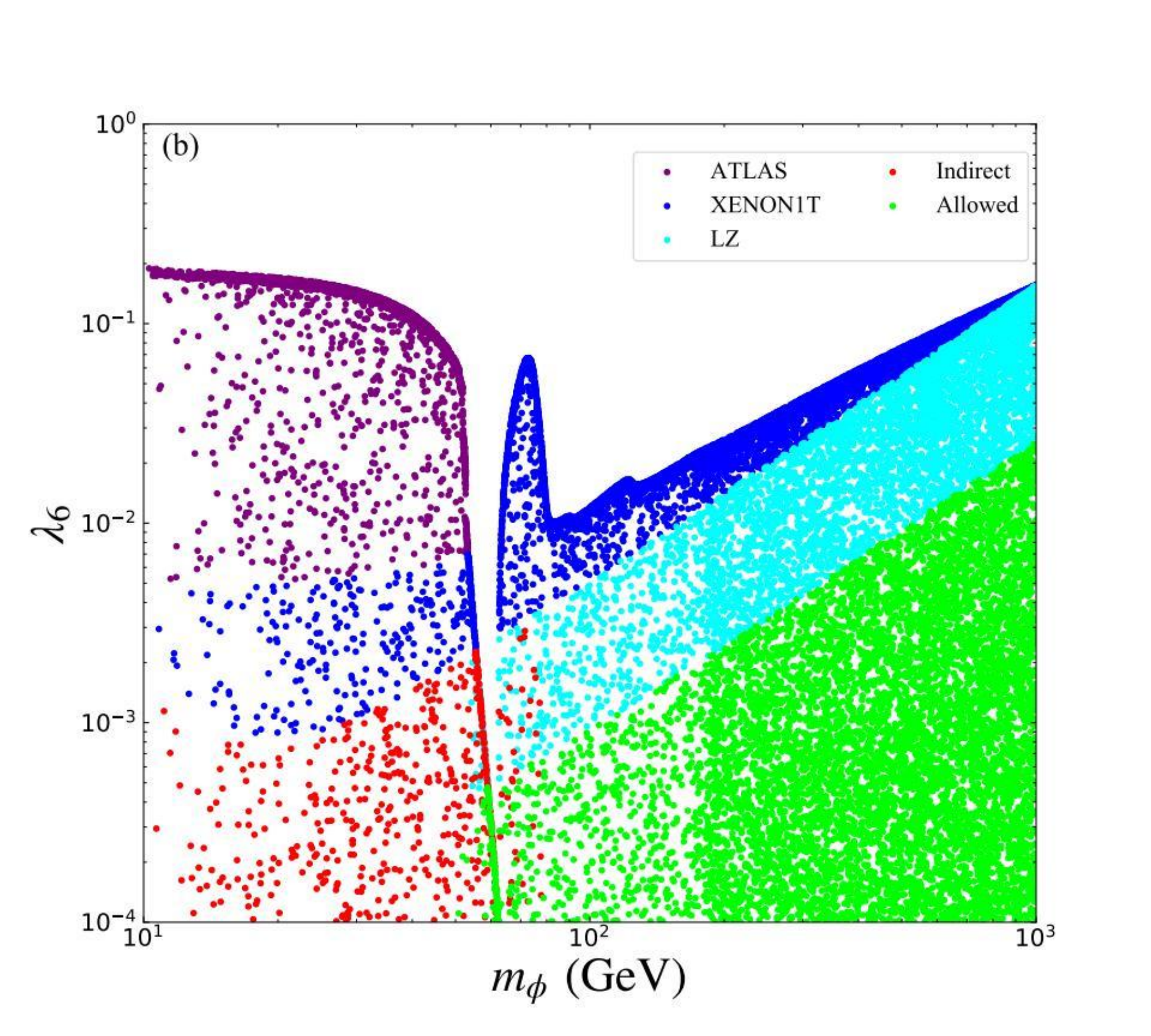}
		\includegraphics[width=0.45\linewidth]{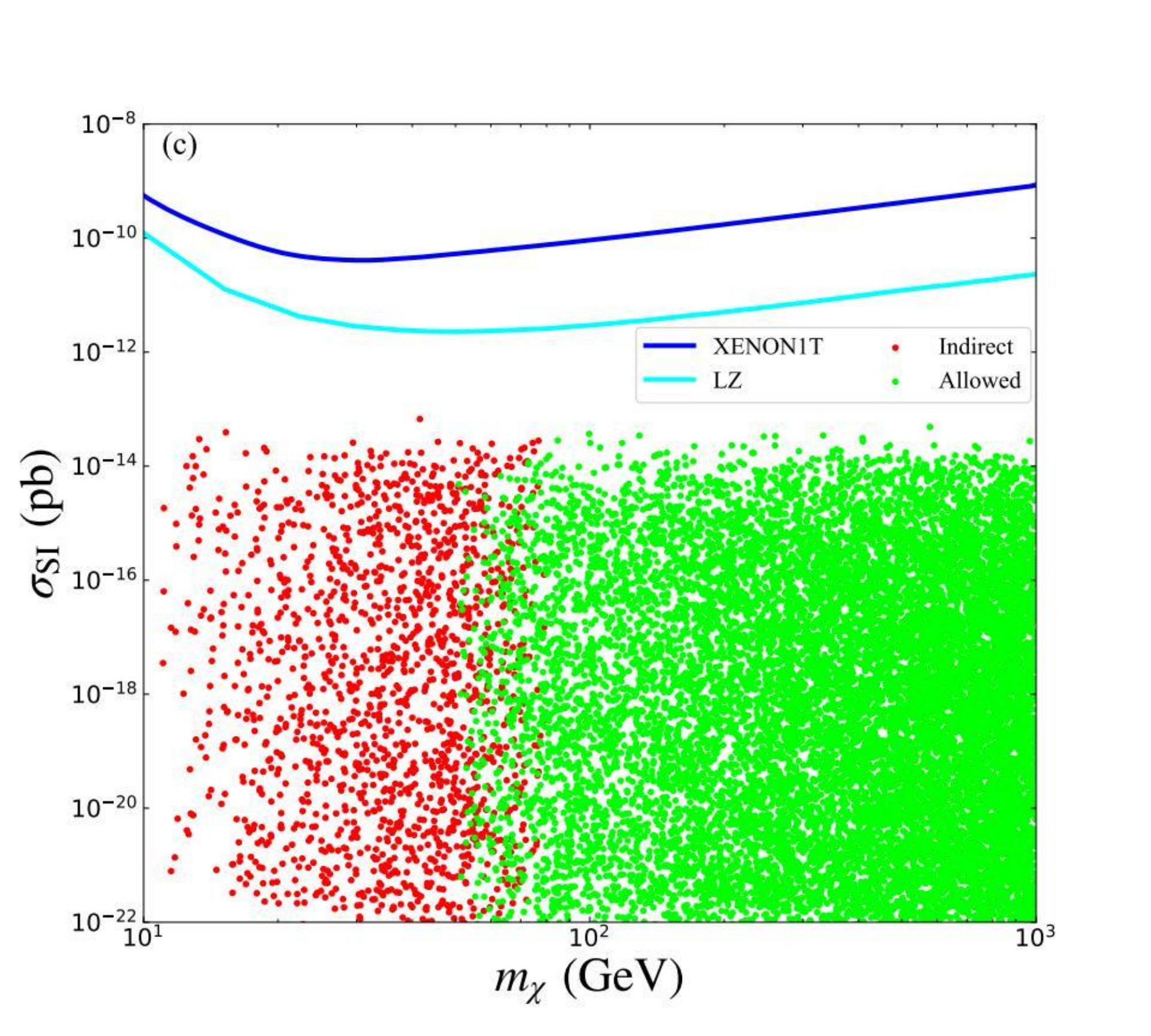}
		\includegraphics[width=0.45\linewidth]{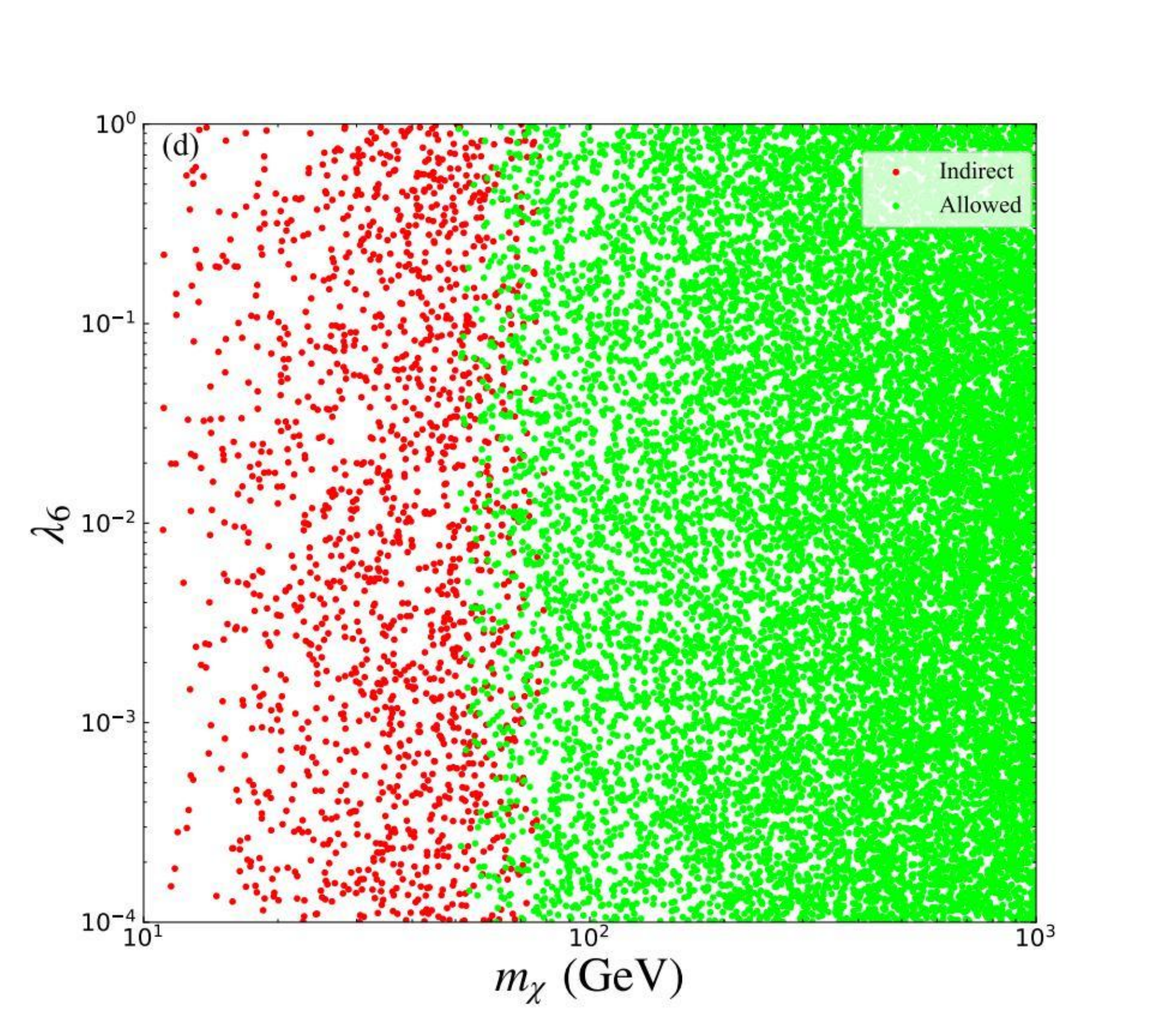}
	\end{center}
	\caption{Constraints from the direct detection experiments. Labels are the same as Fig.~\ref{FIG:HigInv}.}
	\label{FIG:WDir}
\end{figure}

The spin-independent scattering cross sections as a function of DM mass are shown in Fig.~\ref{FIG:WDir}. Since the Higgs portal coupling $\lambda_{6}$ plays a vital important role in direct detection, we also show the parameter space in the $\lambda_6-m_\text{DM}$ plane. Currently, the XENON1T experiment has excluded the samples with a cross section larger than $4\times10^{-11}$ pb for $m_\text{DM}\sim30$ GeV. The future LZ experiment can push this limit down to about $10^{-12}$ pb. For the scalar DM $\phi$, the parameter space with $\lambda_6\gtrsim 0.001$ has been excluded by XENON1T. Combined with the relative contribution in Fig.~\ref{FIG:WRelic}, it is clear that the SM Higgs portal channel could be the dominant one only at the narrow resonance region $m_\phi\sim m_h/2$ or the high mass region $m_\phi\sim1000$ GeV at present. However, this high mass region could be further excluded when there is no signal at the future LZ experiment. This indicates that for $m_\phi\lesssim m_{\Phi_\nu}=200$ GeV, the sterile neutrino portal channel $\phi\phi\to NN$ is usually the dominant one when we take into account the present XENON1T constraint. Meanwhile, $\phi\phi\to NN,\Phi^\dag_\nu \Phi_\nu$ are the dominant channels for $m_{\phi}\gtrsim m_{\Phi_\nu}$ under LZ projected limit. For the fermion DM $\chi$, the effective Higgs portal coupling $\lambda_{h\chi}$ is suppressed by the loop factor, leading to a natural suppression of the  scattering cross section. The predicted value $\sigma_\text{SI}\lesssim10^{-13}$ pb is beyond the reach of XENON1T and LZ experiments, therefore no sample is excluded by direct detection for fermion DM.

\subsection{Indirect Detection}

One distinct property of the sterile neutrino portal DM is that the present day annihilation of DM into $NN$ pair will lead to observable signature at indirect detection experiments. In this work, we consider the indirect detection constraints from the antiproton observations of AMS-02 \cite{AMS:2016oqu} and gamma-rays observations of Fermi-LAT~\cite{Fermi-LAT:2015att}. These experiments are able to probe the DM mass below about 100 GeV \cite{Escudero:2016ksa}. Although there are three possible annihilation channels for scalar DM $\phi$ in our model, the sterile neutrino portal $\phi\phi\to NN$ is the dominant one for $m_\phi\lesssim 100$ GeV under the XENON1T constraint. 

\begin{figure}
	\begin{center}
		\includegraphics[width=0.45\linewidth]{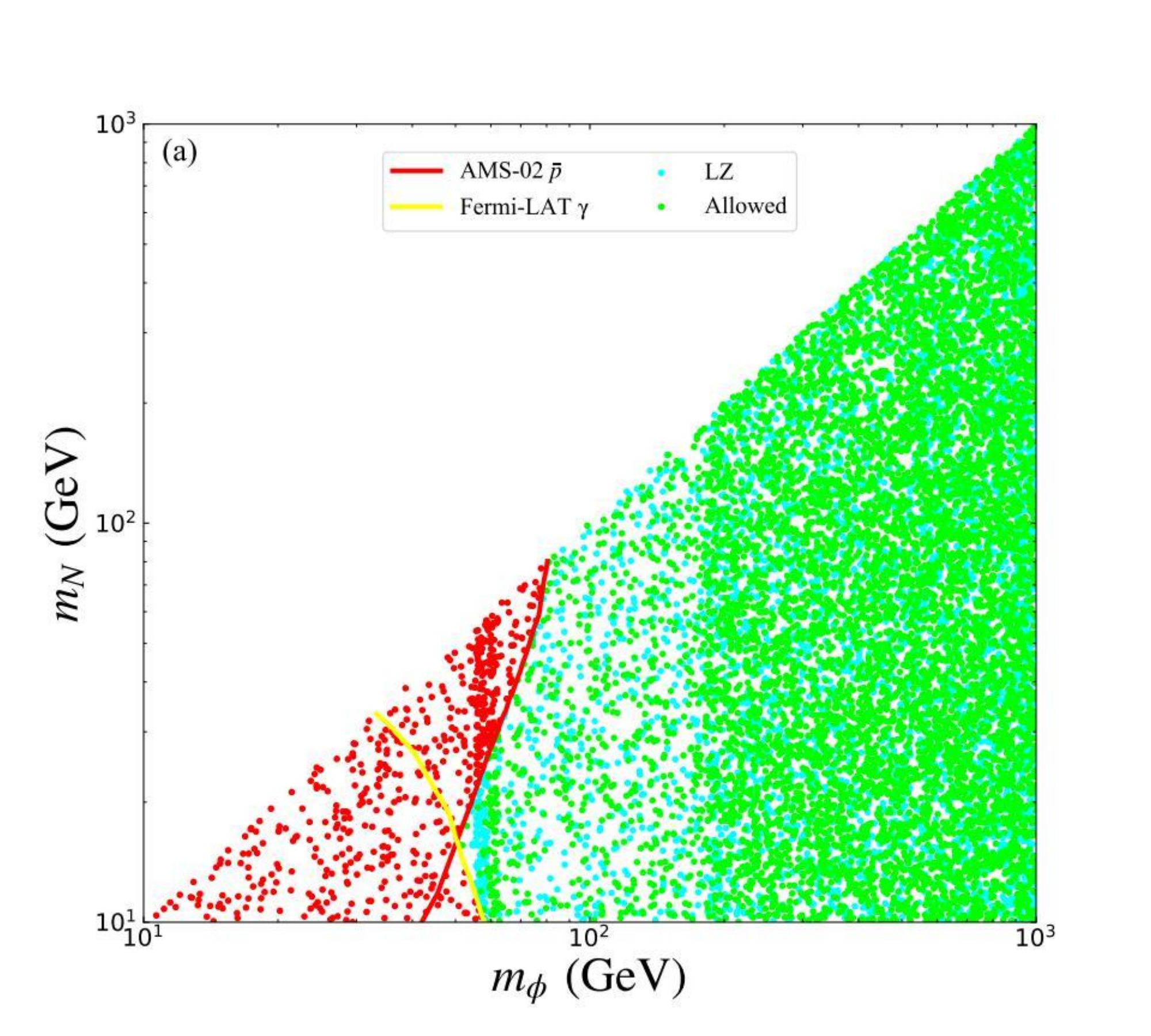}
		\includegraphics[width=0.45\linewidth]{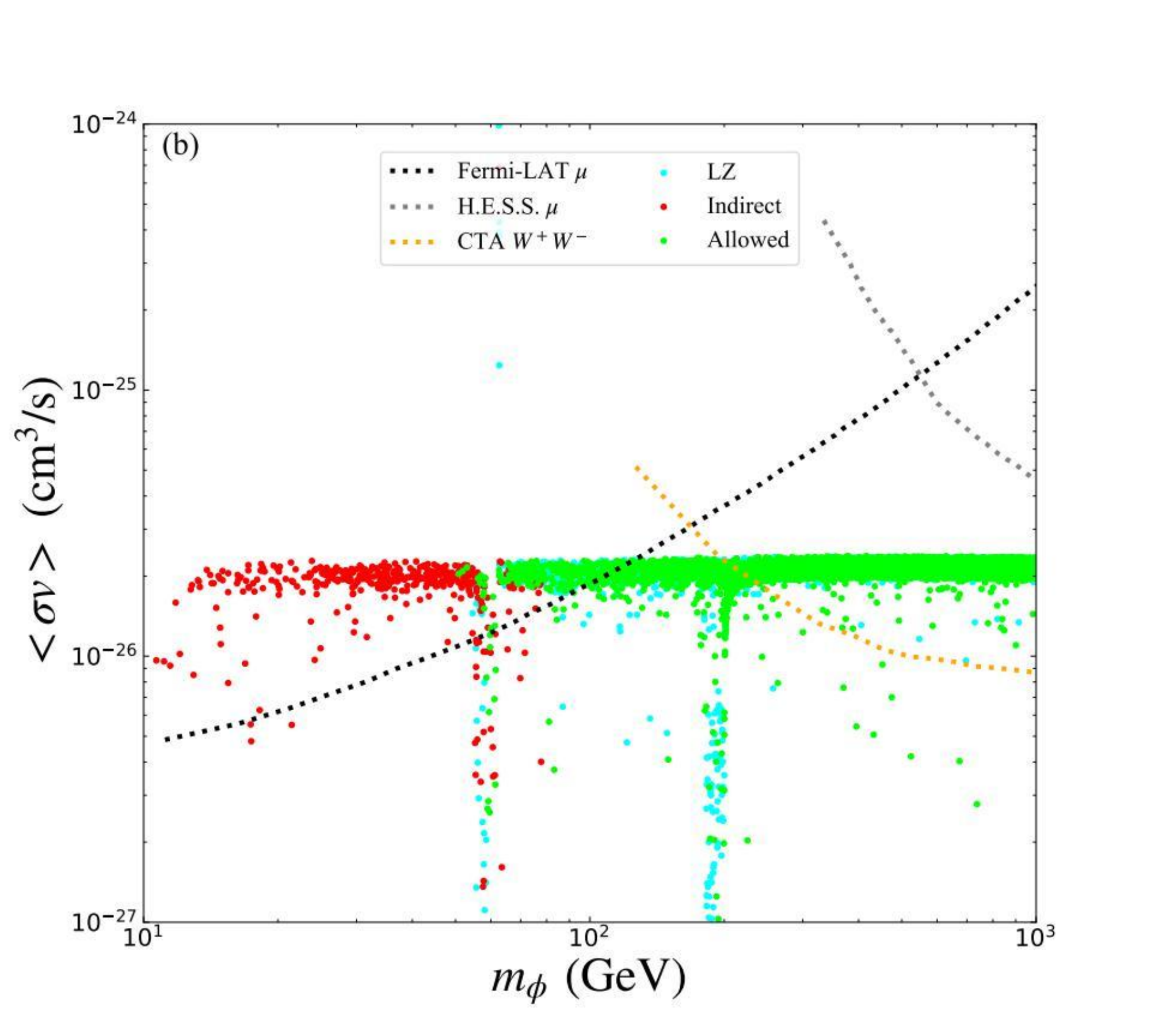}
		\includegraphics[width=0.45\linewidth]{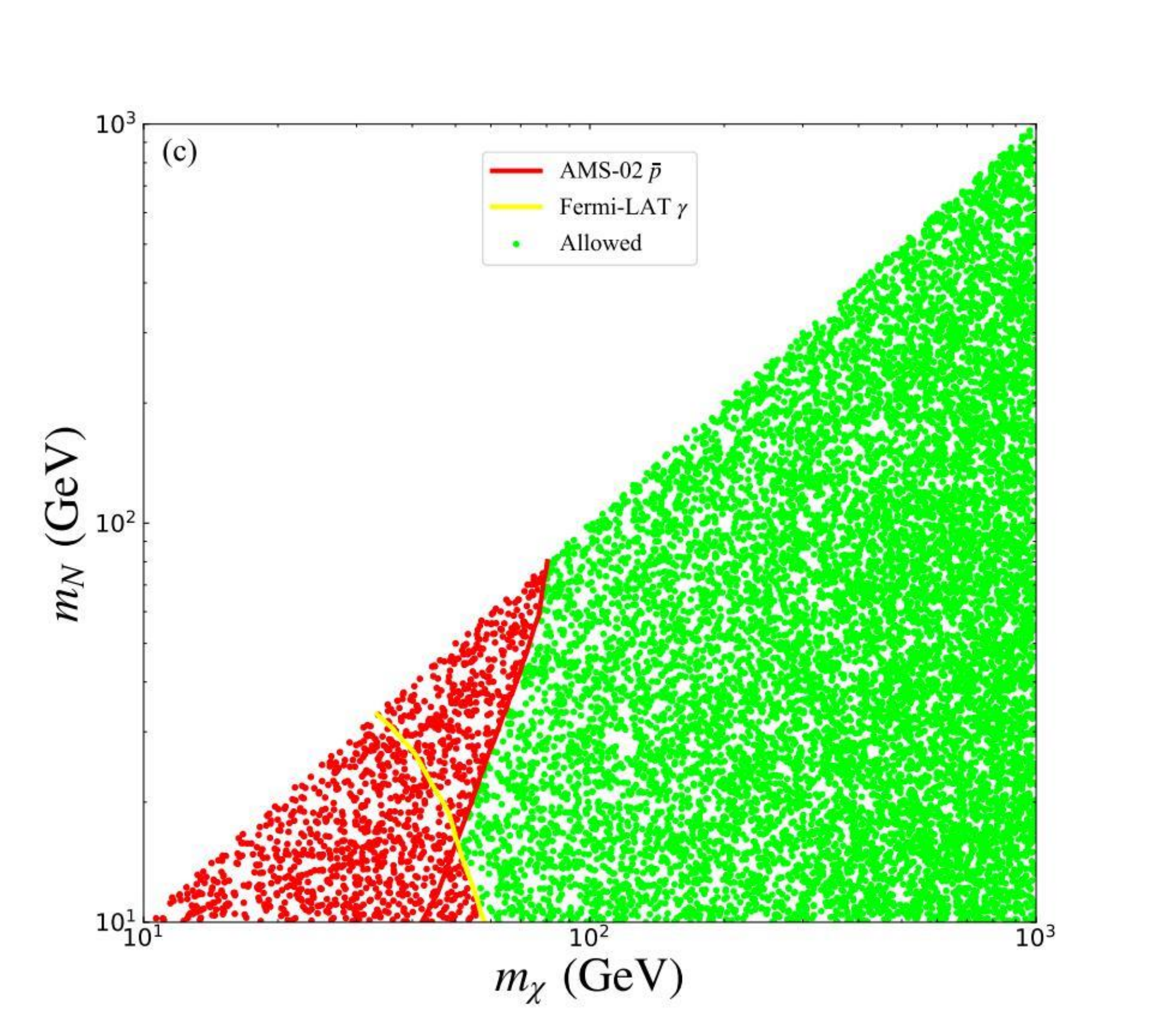}
		\includegraphics[width=0.45\linewidth]{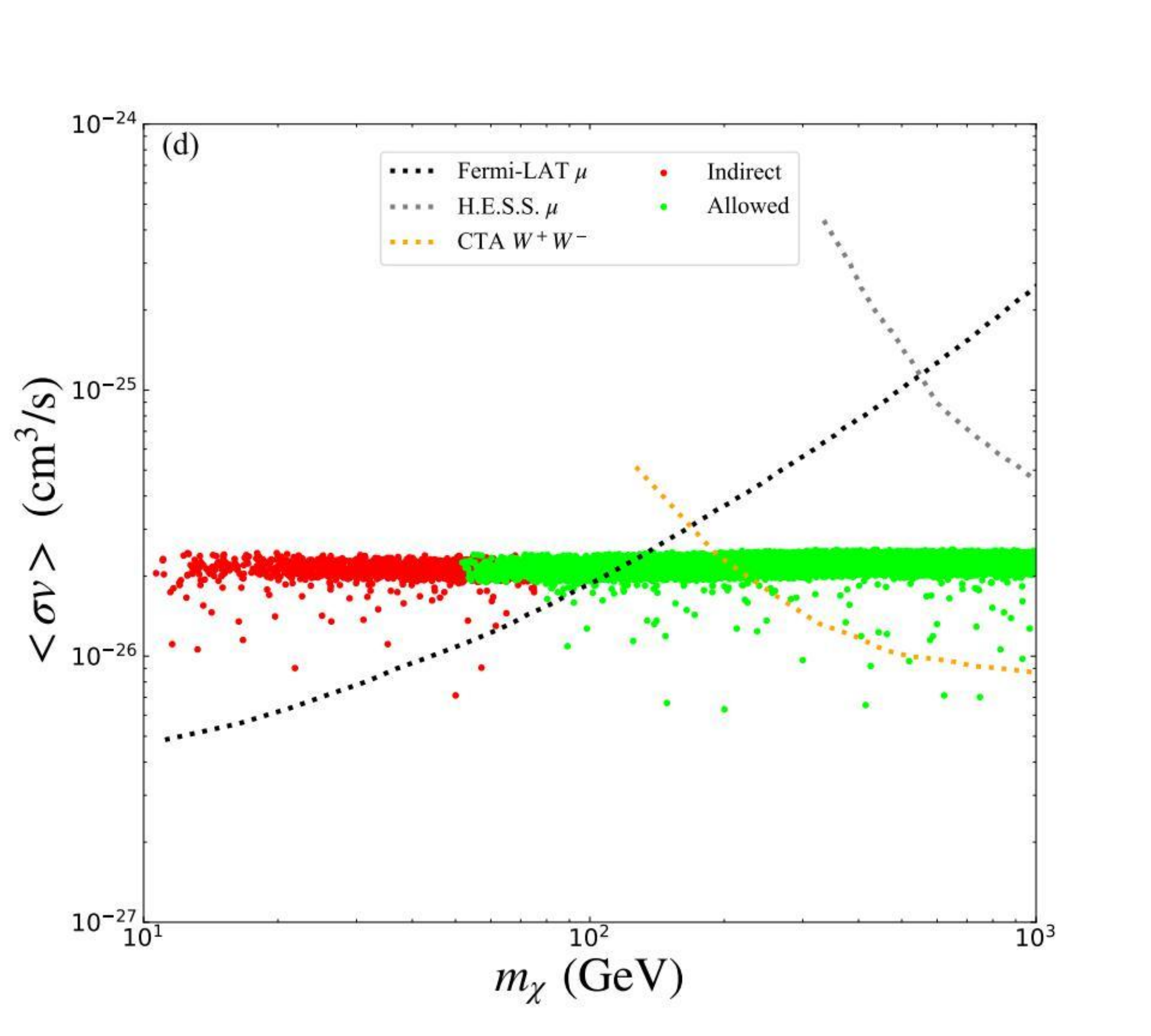}
	\end{center}
	\caption{Constraints from the indirect detection experiments. In the left panel, the red and yellow curves represent the bounds  of antiproton-to-proton flux ratio from AMS-02 \cite{AMS:2016oqu} and gamma-rays in the Milky Way dSphs from Fermi-LAT \cite{Fermi-LAT:2015att} with $\langle\sigma v\rangle=2.2\times10^{-26}~ \rm cm^3/s$, respectively.  In the right panel, the black and gray dotted lines express the upper limits obtained using the $\gamma$ ray data from Fermi-LAT and H.E.S.S. for the muon right-handed neutrino final state with $m_N=10$ GeV \cite{Campos:2017odj}. The orange dotted curve is upper bound from prospective CTA \cite{CTA:2015yxo} with $W^+W^-$ final state. Other labels are the same as Fig.~\ref{FIG:HigInv} . }
	\label{FIG:WInd}
\end{figure}

The exclusion limits of indirect detection experiments are shown in Fig.~\ref{FIG:WInd}. For both the scalar and fermion DM, the thermal average annihilation cross section of $NN$ channel is about $\langle\sigma v\rangle\simeq2.2\times10^{-26} \rm{cm^3/s}$ for most samples, therefore we simply use the exclusion region in the $m_{\rm DM}-m_N$ plane obtained by Ref.~\cite{Batell:2017rol}. Currently, Fermi-LAT experiment is more sensitive for light $m_N$ with $m_\text{DM}\lesssim60$~GeV been excluded. Meanwhile AMS-02 is more sensitive for heavy $m_N$ with $m_\text{DM}\lesssim80$~GeV been excluded. In this way, the indirect detection experiments have set a more stringent constrain than the direct detection experiments in the low mass region of sterile neutrino portal model. Notably, the observed Galactic Center gamma ray excess \cite{Fermi-LAT:2015sau} might be explained by the parameter set $\langle \sigma v\rangle=3.08\times10^{-26}~\text{cm}^3/s, m_\text{DM}=41.3~\GeV,m_N=22.6$ GeV. However, such result is conflict with the indirect limits. For the high mass region, current H.E.S.S. limit does not exclude any samples. In the future, the CTA experiment is able to probe the region above 200 GeV.

\section{FIMP Dark Matter} \label{SEC:FIMP} 

In the case of FIMP DM, we consider that the neutrinophilic particles $\Phi_\nu$ and right-hand neutrinos $N$ are in thermal equilibrium invariably. Meanwhile, we assume that the DM interacts with tiny couplings and never reaches the equilibrium.

\subsection{Relic Density}
\begin{figure}
	\begin{center}
			\includegraphics[width=0.9\linewidth]{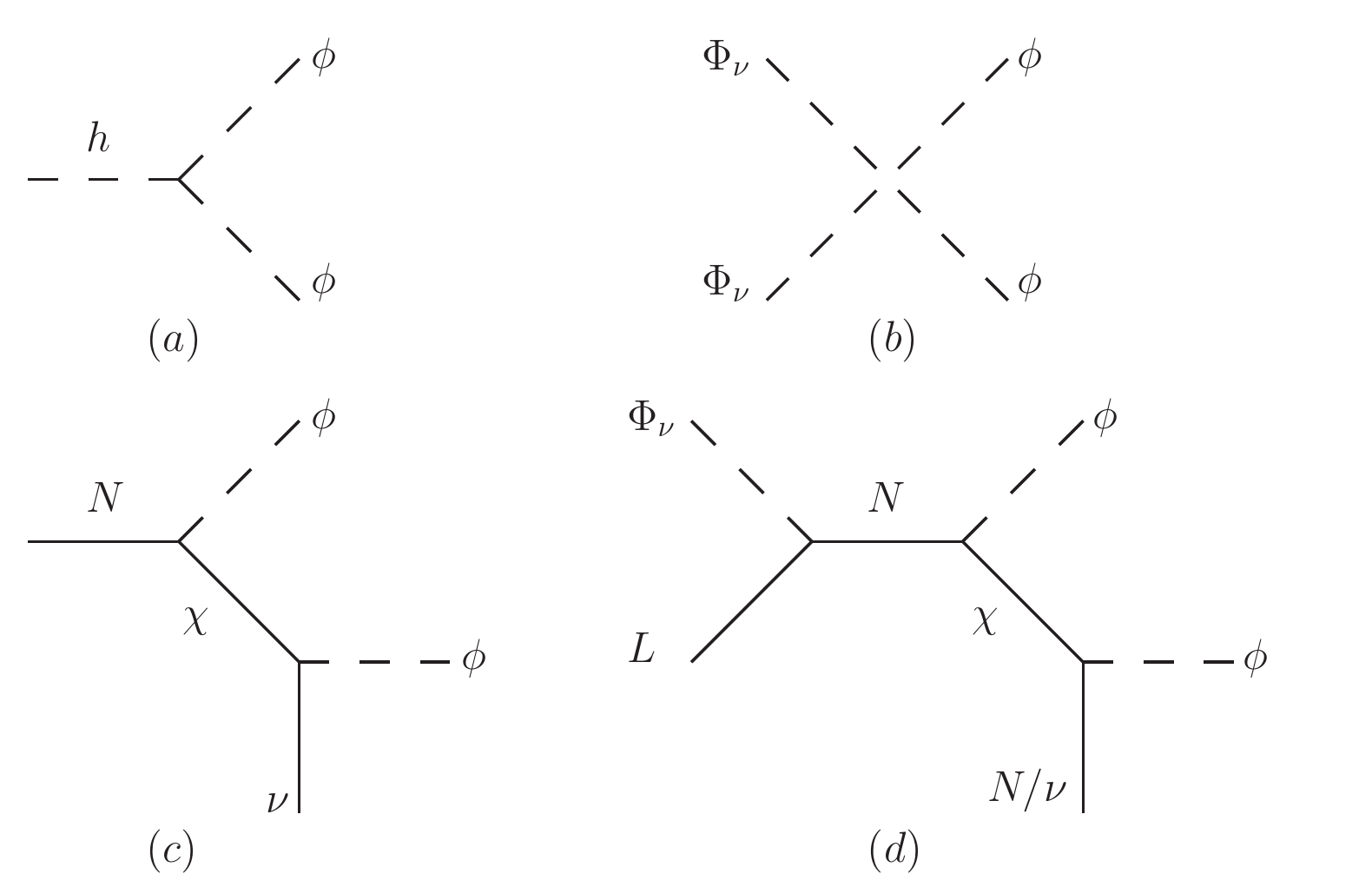}	
	
	\end{center}
	\caption{The relevant Feynman diagrams for scalar FIMP DM $\phi$. }
	\label{FIG:FFeyn}
\end{figure}

If the scalar $\phi$ is FIMP DM, it can be generated from decay or annihilation of SM  Higgs boson $h$ related to $\lambda_6(\phi^\dag \phi)(\Phi^\dag\Phi)$, neutrinophilic scalars connected with $\lambda_7 (\phi^\dag \phi) (\Phi_\nu^\dag\Phi_\nu)$, and sterile neutrinos associated with $\lambda_{ds} \bar{\chi}\phi N$. Due to the small value of $v_\nu$, the scattering process $\Phi_\nu^\dag\Phi_\nu\to\phi \phi$ has the dominant contribution within neutrinophilic portals. The decay of $N\to\phi \chi$ followed by $\chi\to\phi \nu$ is possible when $m_N>m_\phi+m_\chi$, where the $\chi\to \phi\nu$ decay is induced by the  mixing between $N$ and $\nu$.  Otherwise, for a heavier $\chi$, $\phi$ and $\chi$ are associated produced by the annihilation of $\Phi_\nu$ and $L$, which is mediated by the off-shell $N$.  Relevant Feynman diagrams are shown in Fig.~\ref{FIG:FFeyn}.

\begin{figure}
	\begin{center}
			\includegraphics[width=0.9\linewidth]{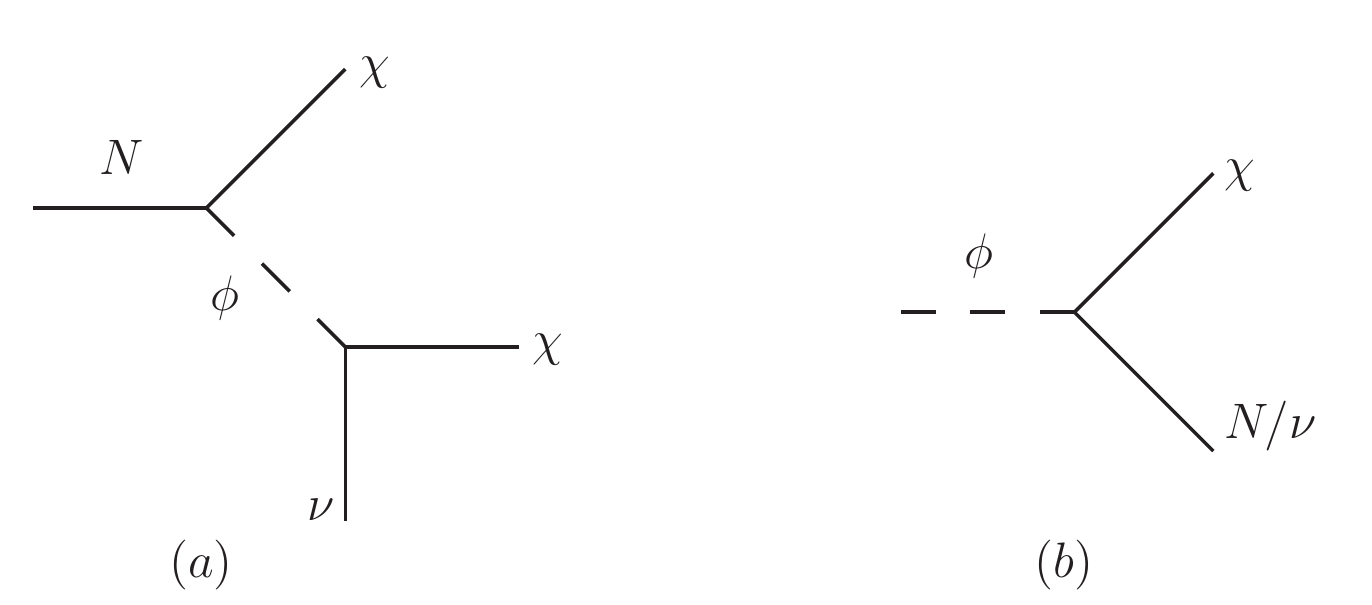}	
	\end{center}
	\caption{The relevant Feynman diagrams for fermion FIMP DM $\chi$.}
	\label{FIG:FFeyn2}
\end{figure}

In the case of fermion FIMP DM $\chi$, it is mainly produced by the decay of $N$ or $\phi$. If $m_N>m_\phi+m_N$, then $\chi$ can be produced by the decay $N\to\chi \phi$ with further decay $\phi\to\chi \nu$. For the opposite case $m_\phi>m_N$,  the decay $\phi\to\chi N$ or $\phi\to\chi \nu$ generates the relic abundance of $\chi$. Relevant Feynman diagrams are shown in Fig.~\ref{FIG:FFeyn2}. The feeble nature of $\chi$ only requires tiny $\lambda_{ds}$, and other couplings of the dark scalar $\phi$, i.e., $\lambda_{6,7}$, are not limited. One scenario is that $\lambda_{6,7}$ is large enough to keep $\phi$ in thermal equilibrium, and the other scenario is that $\phi$ is also feeble with tiny $\lambda_{6,7}$. 

The evolution of abundances of the dark scalar $\phi$ and dark fermion $\chi$  are determined by the following Boltzmann equations
\begin{eqnarray}\label{f-y}
	\frac{dY_\phi}{dz} &= & k^{\star}z\tilde{\Gamma}_{h\to\phi\phi}\left(Y_h^\eq-\frac{Y_h^\eq}{(Y_\phi^\eq)^2}Y_\phi^2\right) 
	+ \frac{k}{z^2} \left \langle \sigma v \right \rangle_{\Phi_\nu\Phi_\nu\to\phi\phi}\left((Y_{\Phi_\nu}^\eq)^2-\left(\frac{Y_{\Phi_\nu}^\eq}{Y_{\phi}^\eq}\right)^2 Y_{\phi}^{2}\right)  \\
	&+&k^{\star}z\tilde{\Gamma}_{N\to\phi\chi}\left(Y_{N}^\eq-\frac{Y_{N}^\eq}{Y_{\phi}^\eq Y_{\chi}^\eq}Y_\phi Y_\chi\right)+\frac{k}{z^2} \left \langle \sigma v \right \rangle_{\Phi_\nu L\to \phi\chi}\left(Y_{\Phi\nu}^{\eq}Y_L^{\eq}-\left(\frac{Y_{\Phi\nu}^\eq Y_L^\eq}{Y_{\phi}^\eq Y_{\chi}^\eq} \right)Y_{\phi}Y_{\chi}\right)
	\nonumber \\
	&+&k^{\star}z\tilde{\Gamma}_{\chi\to\phi N,\phi\nu}\left(Y_{\chi}-\frac{Y_{\chi}^\eq}{Y_{\phi}^\eq}Y_\phi\right)+\frac{k}{z^2} \left \langle \sigma v \right \rangle_{\SM,\SM\to\phi\phi}\left((Y_{\SM}^\eq)^2-\left(\frac{Y_{\SM}^\eq}{Y_{\phi}^\eq}\right)^2 Y_{\phi}^{2}\right)
	\nonumber \\
	&-& \frac{k}{z^2} \left \langle \sigma v \right \rangle_{\phi\phi\to\SM,\Phi_\nu\Phi_\nu}\left(Y_{\phi}^2-(Y_{\phi}^\eq)^2\right) -k^{\star}z \tilde{\Gamma}_{\phi\to\chi N,\chi \nu}\left(Y_{\phi}-\frac{Y_{\phi}^\eq}{Y_{\chi}^\eq}Y_\chi\right),
	\nonumber \\
	\frac{dY_\chi}{dz} &= & k^{\star}z\tilde{\Gamma}_{N\to\phi\chi}\left(Y_{N}^\eq-\frac{Y_{N}^\eq}{Y_{\phi}^\eq Y_{\chi}^\eq}Y_\phi Y_\chi\right)+k^{\star}z \tilde{\Gamma}_{\phi\to\chi N,\chi \nu}\left(Y_{\phi}-\frac{Y_{\phi}^\eq}{Y_{\chi}^\eq}Y_\chi\right)
	 \\ \nonumber
	&+& \frac{k}{z^2} \left \langle \sigma v \right \rangle_{\Phi_\nu L\to\phi\chi}\left(Y_{\Phi_\nu}^{\eq}Y_L^{\eq}-\left(\frac{Y_{\Phi_\nu}^\eq Y_L^\eq}{Y_{\phi}^\eq Y_{\chi}^\eq} \right)Y_{\phi}Y_{\chi}\right)-k^{\star}z\tilde{\Gamma}_{\chi\to\phi N,\phi\nu}\left(Y_{\chi}-\frac{Y_{\chi}^\eq}{Y_{\phi}^\eq}Y_\phi\right),	
\end{eqnarray}
where the parameter $k^{\star}$ satisfies
$k^{*}=\sqrt{45/4\pi^{3}g_{\star}}M_{Pl}/m_\text{DM}^2$.  We use micrOMEGAs \cite{Belanger:2013oya} to numerically calculate the thermal average cross sections $\langle \sigma v \rangle$. The thermal decay width $\tilde{\Gamma}_X$ is defined as $\Gamma_X \mathcal{K}_1/\mathcal{K}_2$ with $\mathcal{K}_{1,2}$ being the first and second modified Bessel Function of the second kind. 
Corresponding decay widths are given by 
 \begin{eqnarray}
	\Gamma_{{N\rightarrow}{\phi\chi}}&=&\frac {\lambda_{ds}^{2}}{16\pi{m_N}}{\left(\frac{(m_N+m_\chi)^2-m_\phi^2}{m_N^2}\right)}{\lambda^{1/2}({m_N}^2,{m_\phi}^2,{m_\chi}^2)}, \nonumber \\
		\Gamma_{\chi\to \phi N} &=& \frac {\lambda_{ds}^{2}}{16\pi{m_\chi}}{\left(\frac{(m_N+m_\chi)^2-m_\phi^2}{m_\chi^2}\right)}{\lambda^{1/2}({m_N}^2,{m_\phi}^2,{m_\chi}^2)},\nonumber \\
	\Gamma_{\chi\to \phi \nu} &=&\frac {\lambda_{ds}^{2}\sin^2\theta\, m_\chi}{16\pi}{\left(\frac{\left(m_\chi^2-m_\phi^2\right)^2}{m_\chi^2}\right)} ,\nonumber \\
	\Gamma_{\phi\to\chi N} &= & \frac {\lambda_{ds}^{2}}{8\pi{m_\phi}}{\left(\frac{m_\phi^2-(m_N+m_\chi)^2}{m_\phi^2}\right)}{\lambda^{1/2}({m_N}^2,{m_\phi}^2,{m_\chi}^2)},\nonumber \\
	\Gamma_{\phi\to \chi \nu} &=& \frac {\lambda_{ds}^{2}\sin^2\theta\, m_\phi}{8\pi}{\left(\frac{\left(m_\phi^2-m_\chi^2\right)^2}{m_\phi^2}\right)},
\end{eqnarray}
The kinematic function $\lambda(a,b,c)$ is
\begin{equation} 
	\lambda(a,b,c)=a^2+b^2+c^2-2ab-2ac-2bc.
\end{equation}
Here the mixing angle $\theta$ can be simply expressed as $\theta=y v_\nu/\sqrt{2}m_N$.  Typically, mixing angle $\theta=1.4\times 10^{-7}$ is obtained with $y=0.01$, $v_\nu=0.01$ GeV and $m_N=500$ GeV.

\begin{figure}
	\begin{center}
		\includegraphics[width=0.45\linewidth]{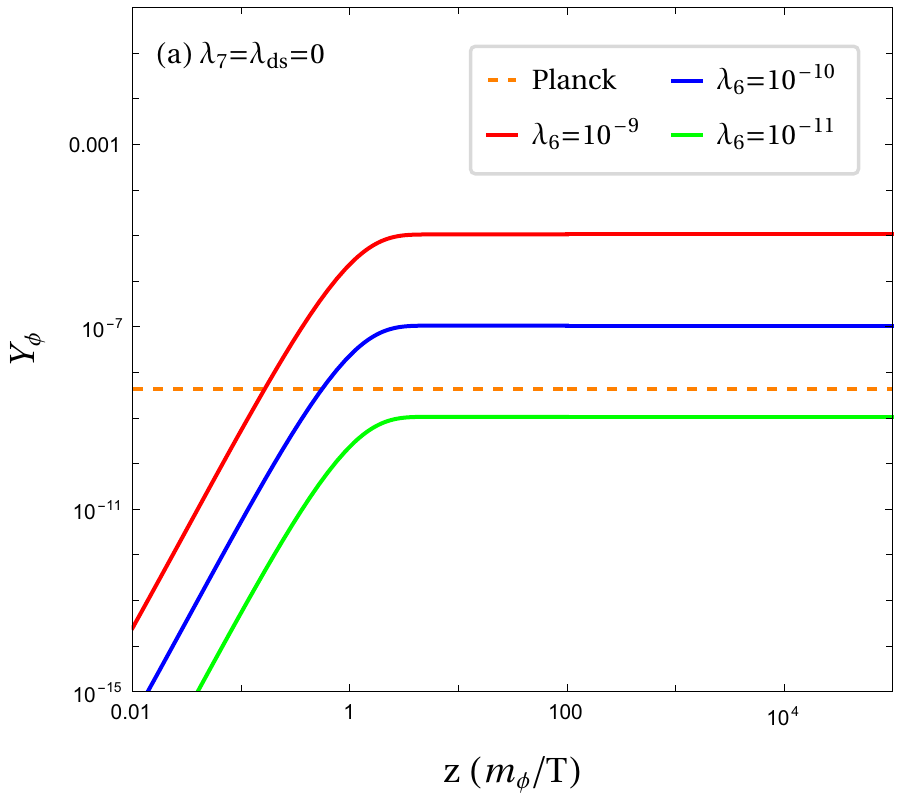}
		\includegraphics[width=0.45\linewidth]{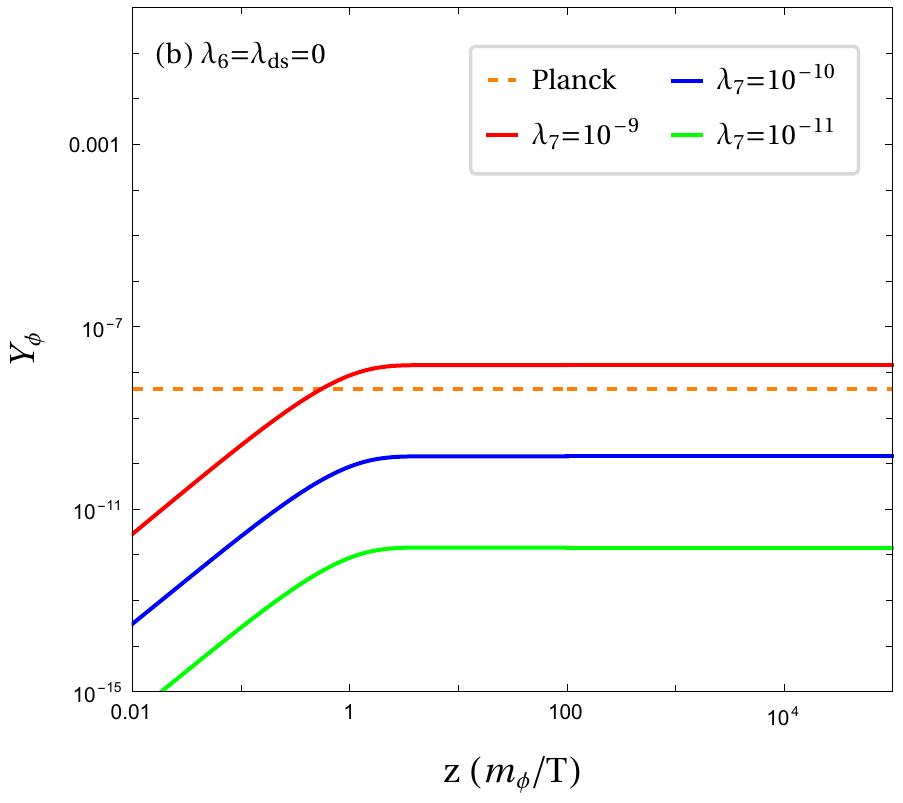}
		\includegraphics[width=0.45\linewidth]{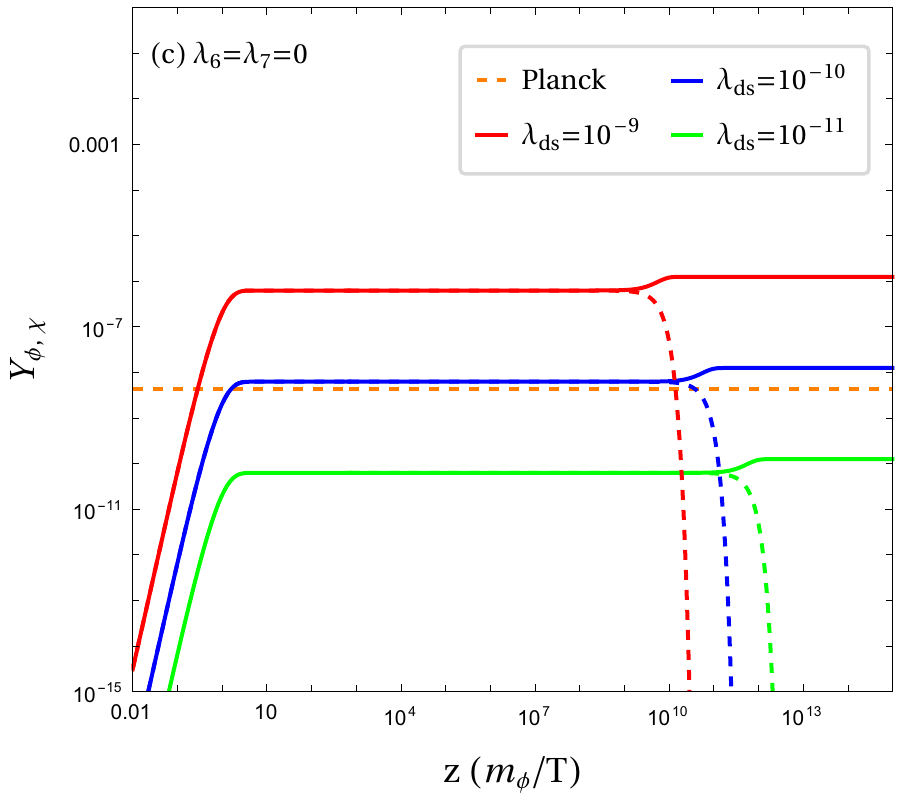}
		\includegraphics[width=0.45\linewidth]{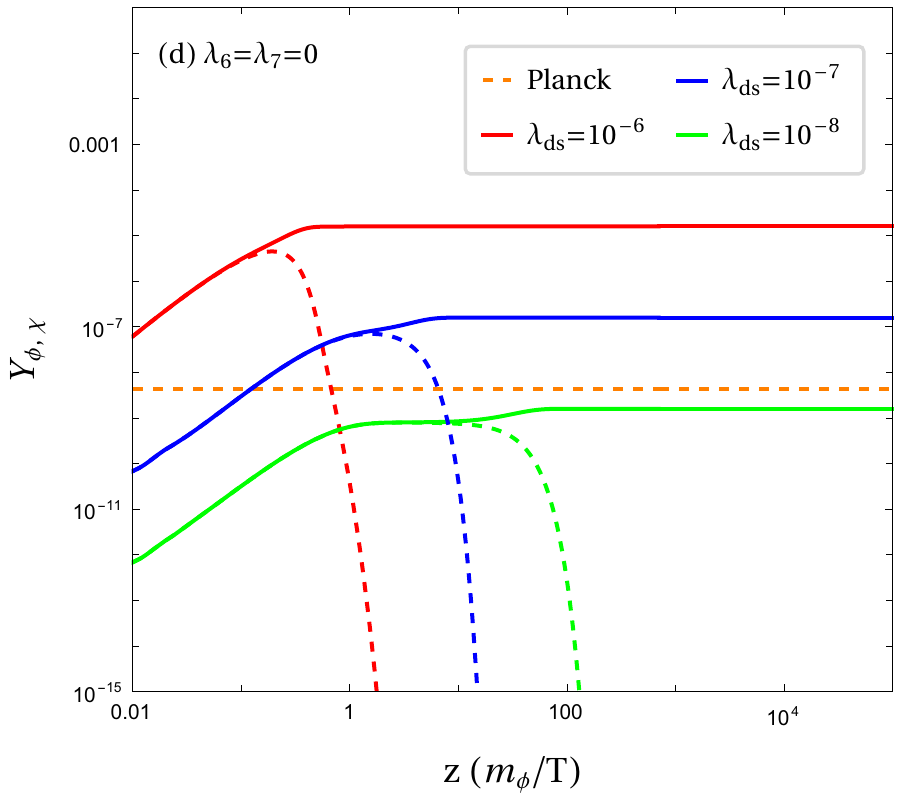}
	\end{center}
	\caption{The evolution of abundance for scalar FIMP $\phi$. The orange horizontal lines correspond to the  Planck observed abundance for $m_\phi=0.1$ GeV. The other dashed lines in subfigure (c) and (d) show the evolution of $Y_\chi$. }
	\label{FIG:FAbun}
\end{figure}

The evolutionary phenomena of scalar FIMP DM is shown in Fig.~\ref{FIG:FAbun}, where we have assume $m_{\phi}=0.1$ GeV. Different from the WIMP scenario, a larger coupling $\lambda_{6,7,ds}$ leads to a larger abundance $Y_\phi$. In panel (a), $\phi$ is mainly generated through the decay $h\to\phi\phi$.  And correct relic density is obtained with $\lambda_{6}\sim\mathcal{O}(10^{-11})$. For the $\Phi_\nu \Phi_\nu\to \phi\phi$ channel, $\lambda_7\sim \mathcal{O}(10^{-9})$ is required to produce $\Omega_\phi h^2\simeq0.12$, which is about two orders of magnitude larger than the coupling $\lambda_6$ for the $h\to \phi\phi$ channel. In Fig.~\ref{FIG:FAbun} (c), both $Y_\phi$ and $Y_\chi$ from the $N\to \phi\chi$ channel are shown, where we assume $m_N=500$ GeV and $m_\chi=10$ GeV. At the very beginning, same amount of $Y_\phi$ and $Y_\chi$ are generated via the $N\to \phi\chi$ decay. Then the late time decay $\chi\to \phi\nu$ translates $Y_\chi$ into $Y_\phi$. This late decay rate is heavily suppressed by the sterile and active neutrino mixing, which might be tested at neutrino experiments \cite{Bandyopadhyay:2020qpn}. In this scenario, the observed abundance is generated with $\lambda_{ds}\sim\mathcal{O}(10^{-10})$. In panel (d), we show the scenario with $m_\chi=150$ GeV, $m_N=100$ GeV and $y=0.01$. Comparing with the pair production channel $NN\to \phi\phi$, the new associated production channel $\Phi_\nu L\to N^*\to \phi \chi$ followed by $\chi\to \phi N$ becomes the dominant one. Correct relic abundance is realized with $\lambda_{ds}\sim\mathcal{O}(10^{-8})$, which is about two orders of magnitudes larger than the scenario of on-shell decay $N\to \phi\chi$. Different from the scenario in panel (c), here the $\chi\to \phi N$ decay is not suppressed by the small mixing angle $\theta$. Therefore, after $\chi$ is associated produced, it quickly decays into $\phi N$.

\begin{figure}
	\begin{center}
		\includegraphics[width=0.45\linewidth]{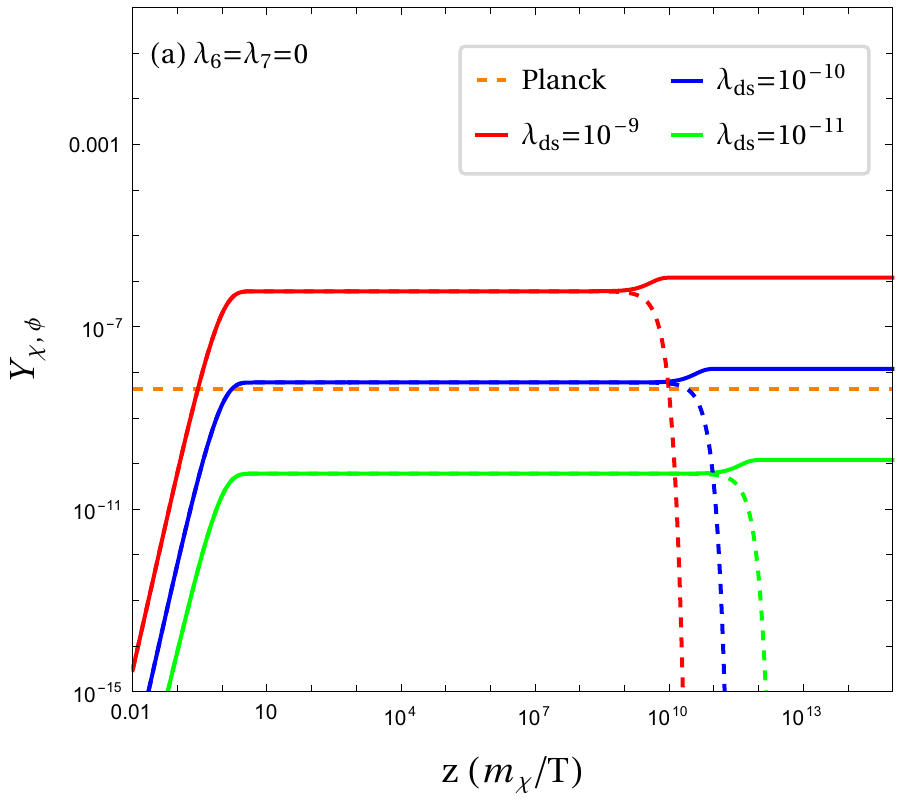}
		\includegraphics[width=0.45\linewidth]{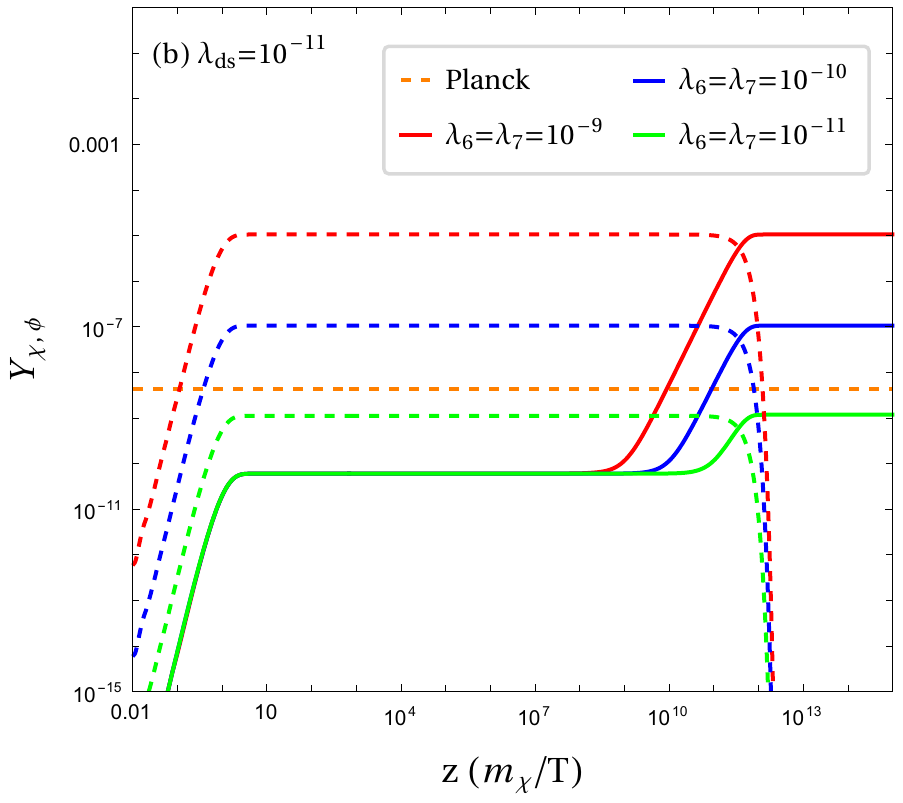}
		\includegraphics[width=0.45\linewidth]{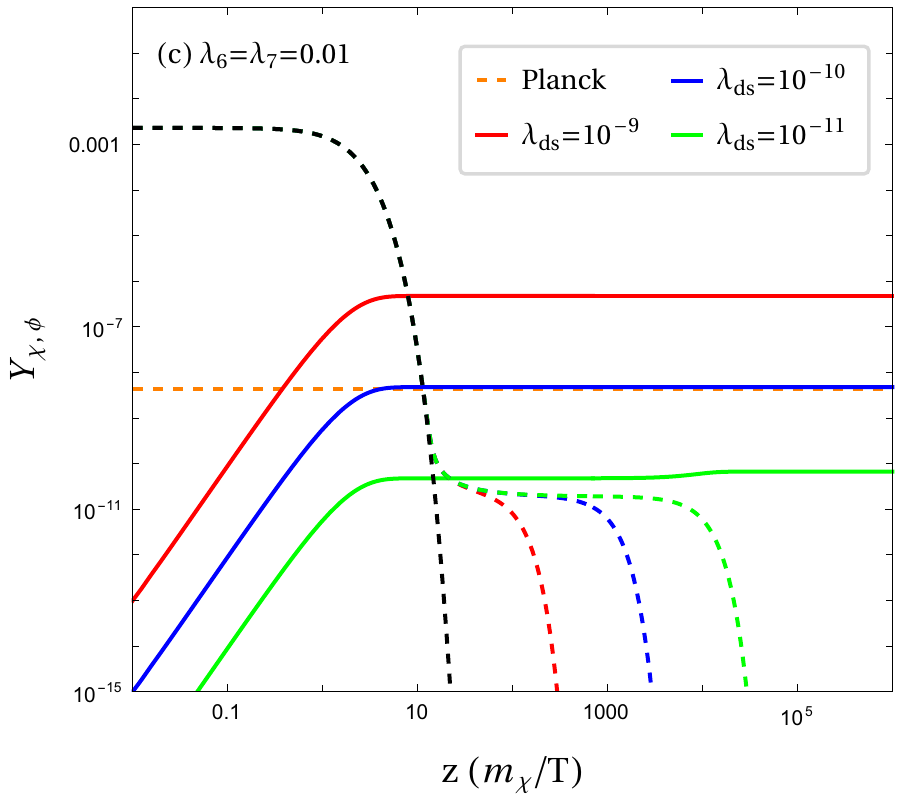}
		\includegraphics[width=0.45\linewidth]{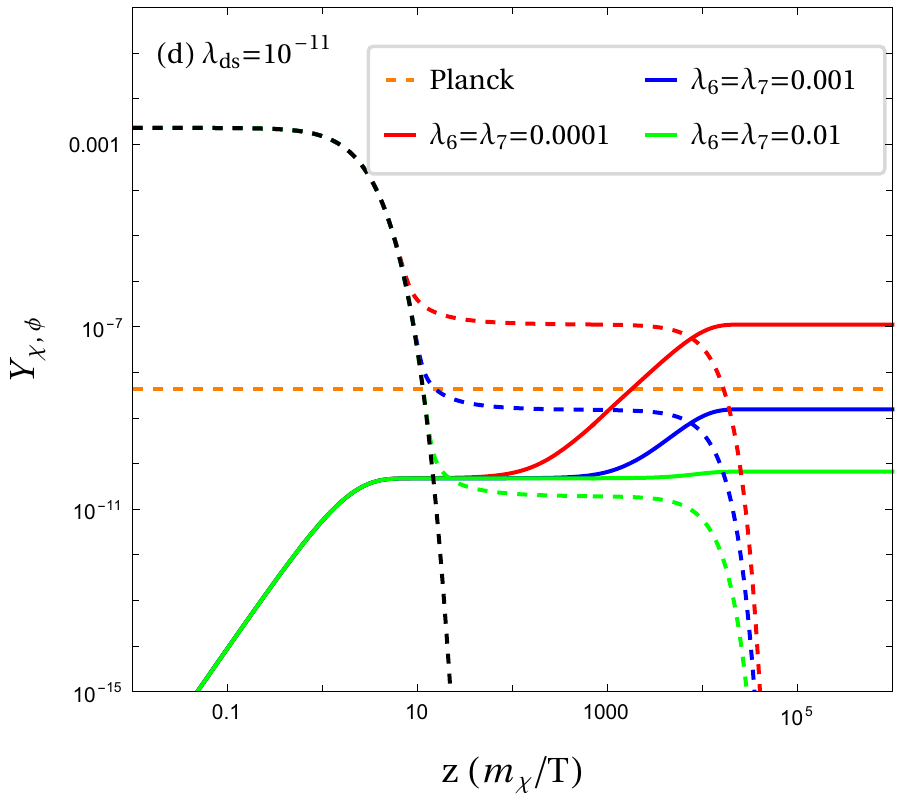}
	\end{center}
	\caption{The evolution of abundance for fermion FIMP $\chi$. The orange horizontal lines correspond to the  Planck observed abundance for $m_\chi=0.1$ GeV. The other dashed lines are the evolution of $Y_\phi$.  }
	\label{FIG:FAbun2}
\end{figure}

The evolution of $Y_\chi$ and $Y_\phi$ for fermion FIMP scenario are shown in Fig.~\ref{FIG:FAbun2}, where we also fix $m_\chi=0.1$~GeV. Panels (a) and (b) correspond to FIMP DM $\chi$ produced by direct decay of $N\to \phi\chi$ with cascade decay $\phi\to \chi\nu$, which is shown in Fig.~\ref{FIG:FFeyn2} (a). During the calculation, we have fixed $m_\chi=0.1$ GeV, $m_N=500$ GeV and $m_\phi=10$ GeV. With vanishing scalar coupling $\lambda_{6,7}=0$, both $\phi$ and $\chi$ are dominant from the decay $N\to \phi \chi$. And the transition $Y_\phi\to Y_\chi$ happens via the late decay $\phi\to \chi \nu$. Correct relic density is obtained with $\lambda_{ds}\sim\mathcal{O}(10^{-10})$ when $N\to \phi\chi$ contributes solely. However, the other scalar interaction $\lambda_{6,7}$-terms of the dark scalar $\phi$ have great impact on the evolution of $Y_\chi$. First, we assume that $\phi$ is also feeble interacting. The result is shown in panel (b) of Fig.~\ref{FIG:FAbun2}. Since the direct decay $h\to \phi\phi$ is kinematically allowed for $m_\phi=10$ GeV, the production of $\phi$ is subdominant by the decay $N\to \phi \chi$ for $\lambda_6=\lambda_7=\lambda_{ds}$. So when $\lambda_6=\lambda_7\gtrsim \lambda_{ds}$, the final abundance $Y_\chi$ is actually controlled by the dark scalar produced from the scalar portal channels. In another case of $m_\phi>m_N$, the generation of DM $\chi$ is mainly determined by the decay of $\phi$.  The coupling $\lambda_{ds}$ only determines the decay rate of $\phi$, and has no effect on the final DM abundance. Evolution of $Y_\phi$ is similar with Fig.~\ref{FIG:FAbun} (a) and (b), but finally $Y_\phi$ is translated into $Y_\chi$ via decay $\phi\to \chi N, \chi \nu$.

Next, let's consider that the quartic coupling $\lambda_{6,7}$ are relative large to keep $\phi$ in thermal equilibrium.  The results are shown in panel (c) and (d) of Fig.~\ref{FIG:FAbun2} with  $m_\chi=0.1$ GeV, $m_\phi=500$ GeV and $m_N=100$~GeV, where the abundance of $\phi$ is controlled by the annihilation process $\phi\phi\to \SM \, \SM,\Phi_\nu\Phi_\nu$ and  $\phi\to \chi N$ is the dominant production mode of fermion DM. For $\lambda_6=\lambda_7=0.01$, the abundance of $\phi$ after freeze-out is smaller than the abundance of $\chi$ from freeze-in with $\lambda_{ds}>10^{-11}$. By decreasing the quartic coupling $\lambda_{6,7}$, the abundance of $\phi$ will increase. The decay of $\phi$ after freeze-out is the dominant contribution when $\lambda_6=\lambda_7\lesssim0.01$ and $\lambda_{ds}=10^{-11}$. As for the case of $m_N>m_\phi$, evolution of $Y_\phi$ is similar with Fig.~\ref{FIG:FAbun2} (c) and (d), but the lifetime of $\phi$ is much longer due to the tiny late decay rate of $\phi\to \chi \nu$.

\begin{figure}
	\begin{center}
		\includegraphics[width=0.45\linewidth]{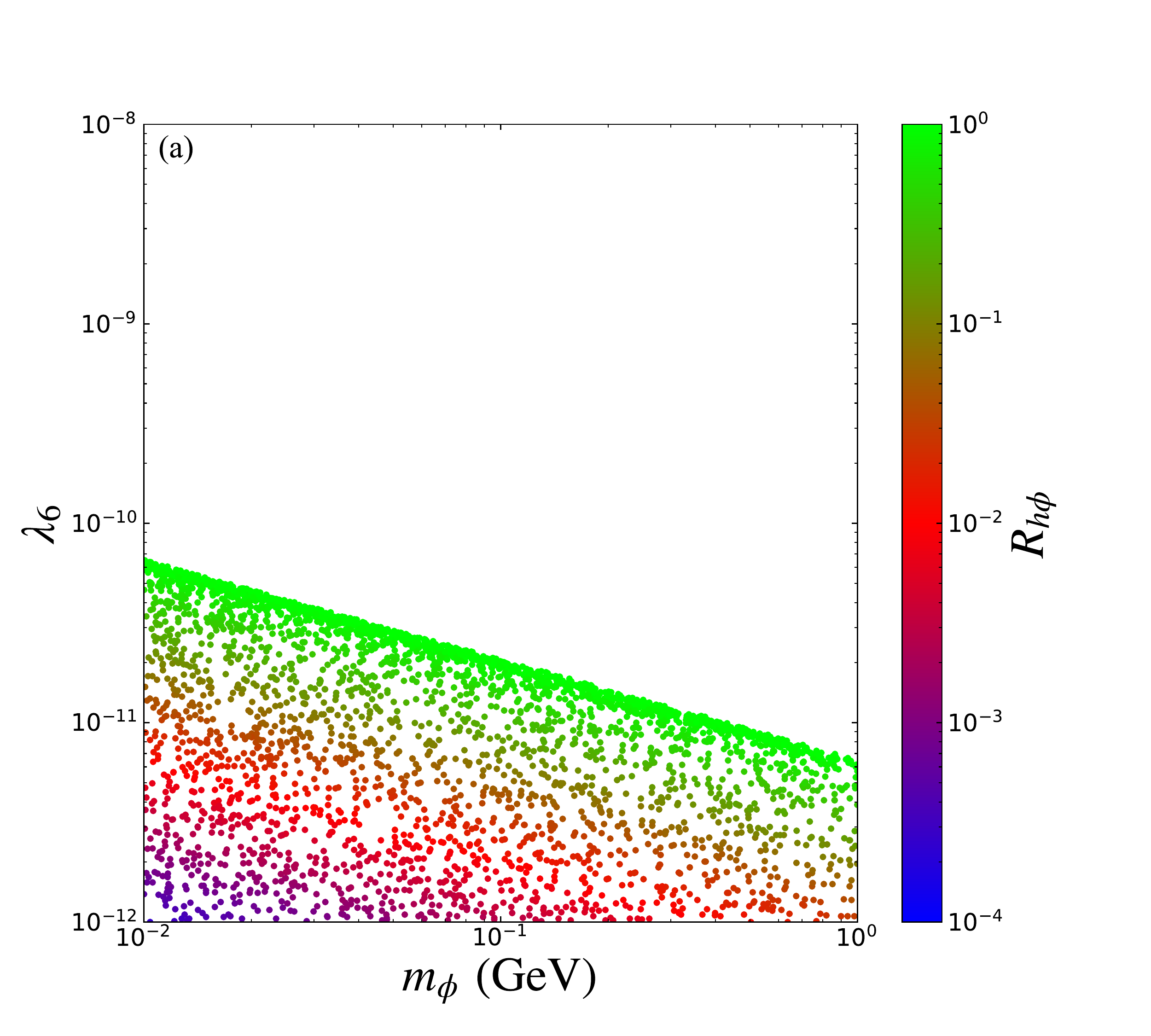}
		\includegraphics[width=0.45\linewidth]{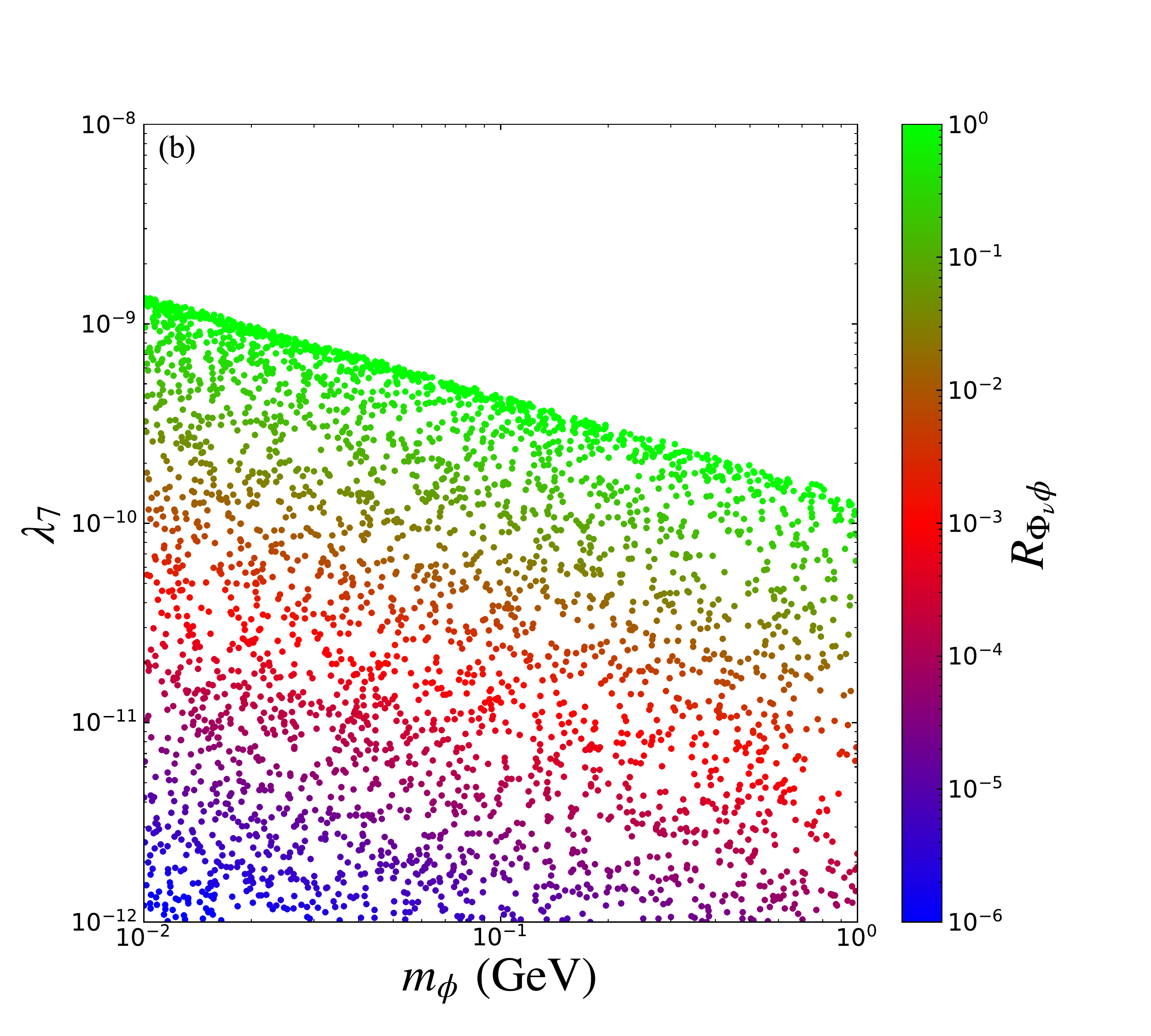}
		\includegraphics[width=0.45\linewidth]{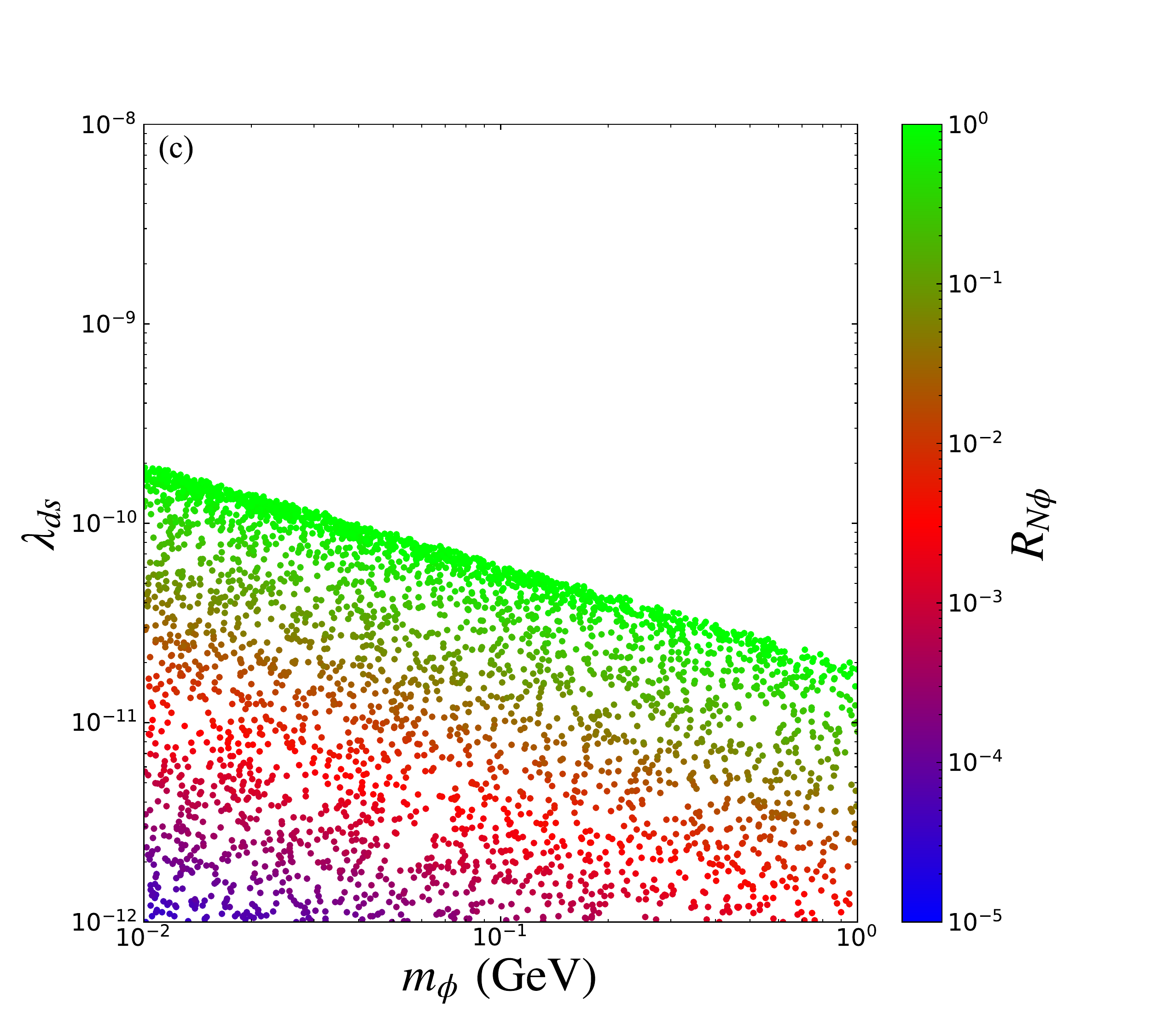}
		\includegraphics[width=0.45\linewidth]{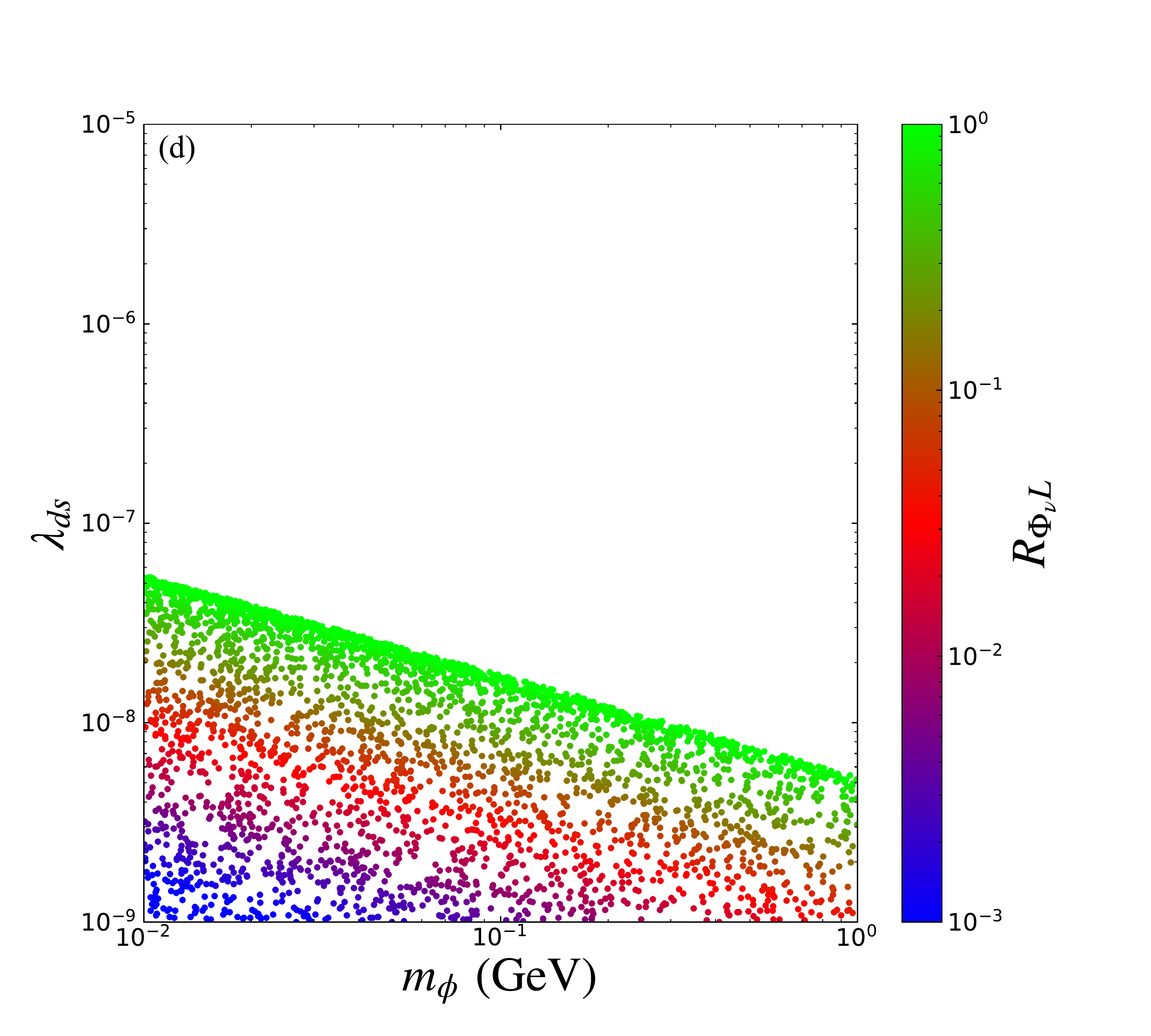}
	\end{center}
	\caption{Allowed parameter space with correct relic density for scalar FIMP DM $\phi$. Panel (a)-(c) correspond to the scenario $m_N>m_\chi$, and panel (d) corresponds to $m_N<m_\chi$.}
	\label{FIG:FRelic}
\end{figure}

\subsection{Scanning Results}

Corresponding to the above benchmark situations, we perform a random scan to obtain the parameter space of FIMP DM within $3\sigma$ range of the observed abundance.  We first consider the scalar DM $\phi$. Depending on the mass relation between $m_N$ and $m_\chi$, the sterile neutrino portal channel requires quite different values of $\lambda_{ds}$. So we divide the scan into two parts. For the scenario with $m_N>m_\chi$, we scan the following parameter space:
\begin{eqnarray}
	\begin{aligned}
		m_\phi\in[0.01,1]~\GeV, m_\chi\in[1,100]~\GeV, \lambda_{6,7,ds}\in[10^{-12}, 10^{-8}].
	\end{aligned}
\end{eqnarray}
During the scan, we fix $m_N=500$ GeV for simplicity. The results are shown in Fig.~\ref{FIG:FRelic} (a)-(c), where the relative contribution of individual channels are defined as
\begin{eqnarray}
	R_{h\phi}=\frac{\Omega_{h\phi}h^2}{\Omega_{\rm DM}h^2},
	R_{\Phi_{\nu}\phi}=\frac{\Omega_{\Phi_{\nu}\phi}h^2}{\Omega_{\rm DM}h^2},	R_{N\phi}=\frac{\Omega_{N\phi}h^2}{\Omega_{\rm DM}h^2}.
\end{eqnarray}
Hence $R_{h\phi}$, $R_{\Phi_{\nu}\phi}$ and $R_{N\phi}$ represent the contribution of $h\to \phi\phi$, $\Phi_{\nu}\Phi_{\nu}\to \phi\phi$ and $N\to \phi \chi$ channel as shown in Fig.~\ref{FIG:FFeyn} (a)-(c), respectively. It is clear that the the DM mass is inversely proportional to the coupling coefficient for the same contribution. The $h\to \phi\phi$, $\Phi_{\nu}\Phi_{\nu}\to \phi\phi$ and $N\to \phi \chi$ channel become the dominant one when the corresponding coefficients individually satisfy $\lambda_6\in[6.5 \times 10^{-12},6.8 \times 10^{-11}]$, $\lambda_7\in[1.3 \times 10^{-11},1.3 \times 10^{-9}]$, $\lambda_{ds}\in[2.0 \times 10^{-11},2.0 \times 10^{-10}]$ with $m_\phi$ decreasing from 1 GeV to 0.01~GeV. 

For the scenario with $m_N<m_\chi$, the dominant contribution of annihilation channel $\Phi_\nu L\to N^*\to \phi \chi$ requires that the coupling $\lambda_{ds}$ is about two to three orders of magnitudes larger than the $m_N>m_\chi$ scenario, so we scan the parameter space
\begin{eqnarray}
	\begin{aligned}
	 m_\phi\in[0.01,1]~\GeV, m_\chi\in[100,200]~\GeV, \lambda_{6,7}\in[10^{-12}, 10^{-8}], \lambda_{ds}\in[10^{-9}, 10^{-5}], 
	\end{aligned}
\end{eqnarray}
where $m_N=100$ GeV and $y=0.01$ are fixed for illustration. $m_\chi$ is not greater than $m_{\Phi_\nu}$ to make the scattering process occur naturally. In addition, we take $m_\phi+m_N<m_\chi$ to ensure the decay channel $\chi\to\phi N$ is kinematically allowed during scanning.  Notably, the relative contributions of $h\to \phi\phi$ and $\Phi_\nu\Phi_\nu\to \phi\phi$ channel are the same as previous scenario with $m_N>m_\chi$, so we only need to show the result of $\Phi_\nu L\to \phi\chi$ channel in Fig.~\ref{FIG:FRelic} (d). We use the notation  $R_{\Phi_{\nu}L}=\Omega_{\Phi_{\nu}L\to\phi\chi}h^2/\Omega_{\rm DM}h^2$  to express the relative contribution of this off-shell channel. The coupling  $\lambda_{ds}\in[5.3\times 10^{-9},5.3\times 10^{-8}]$ will lead this channel to be the dominant one.

\begin{figure}
	\begin{center}
		\includegraphics[width=0.45\linewidth]{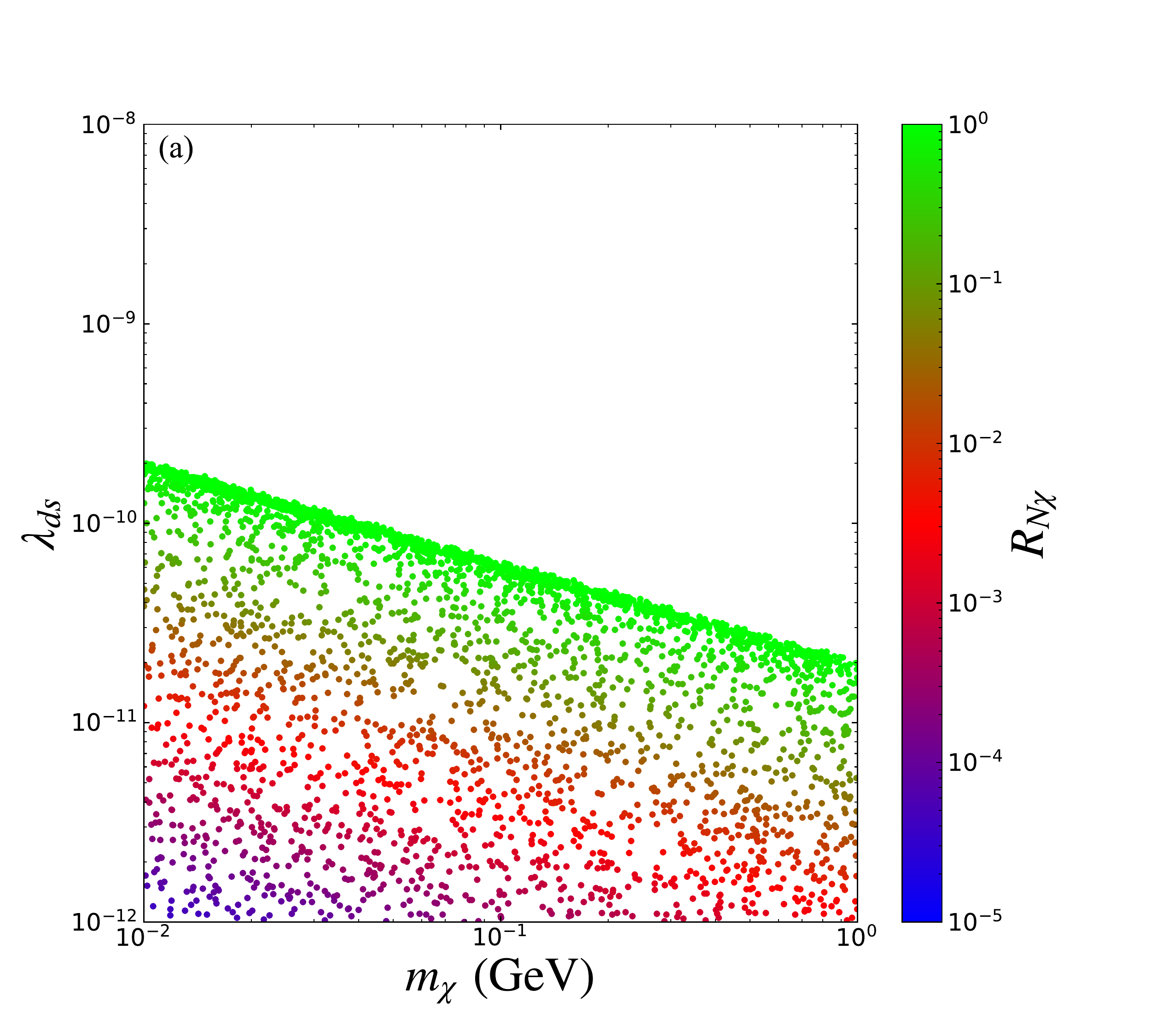}
		\includegraphics[width=0.45\linewidth]{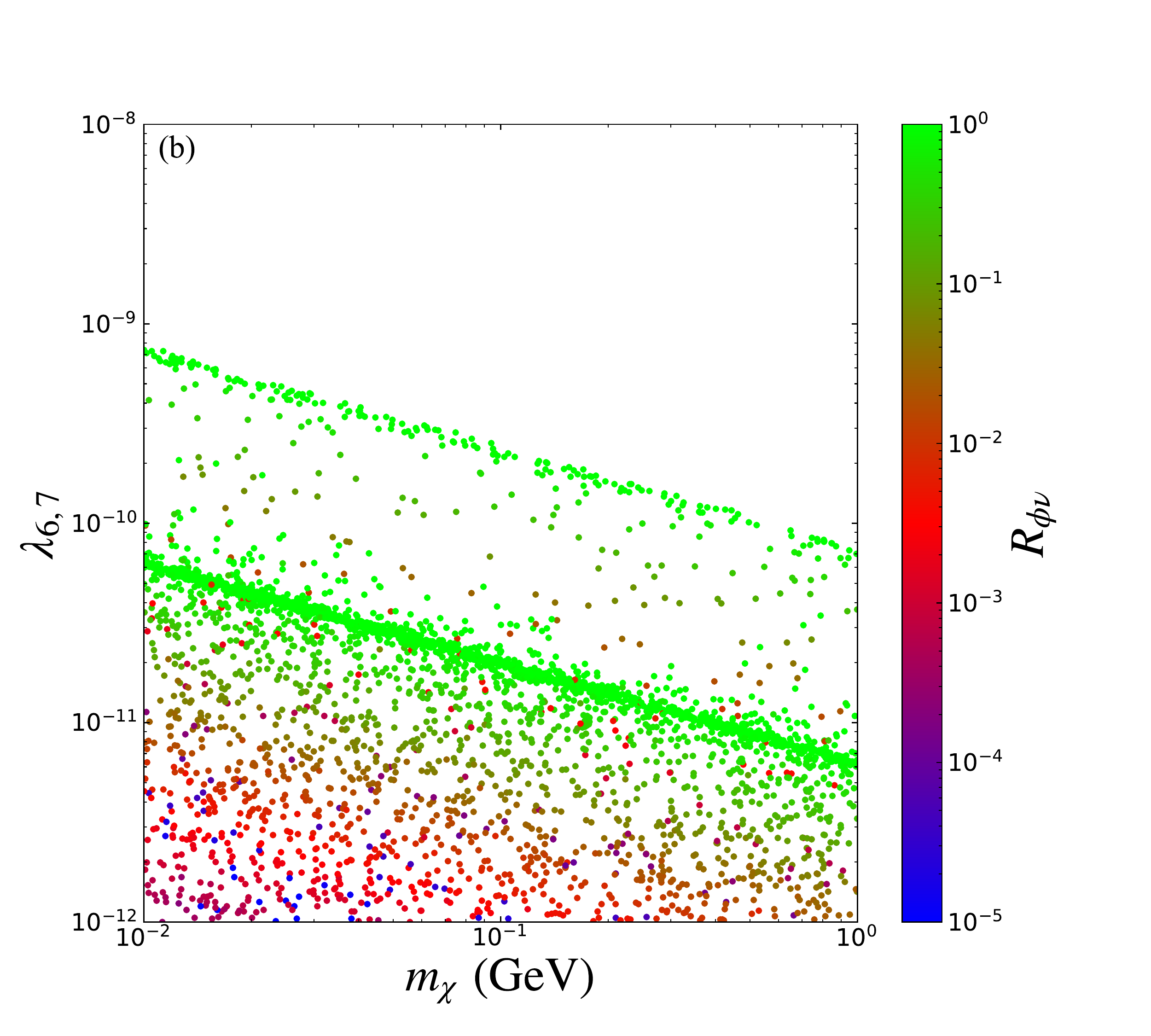}
		\includegraphics[width=0.45\linewidth]{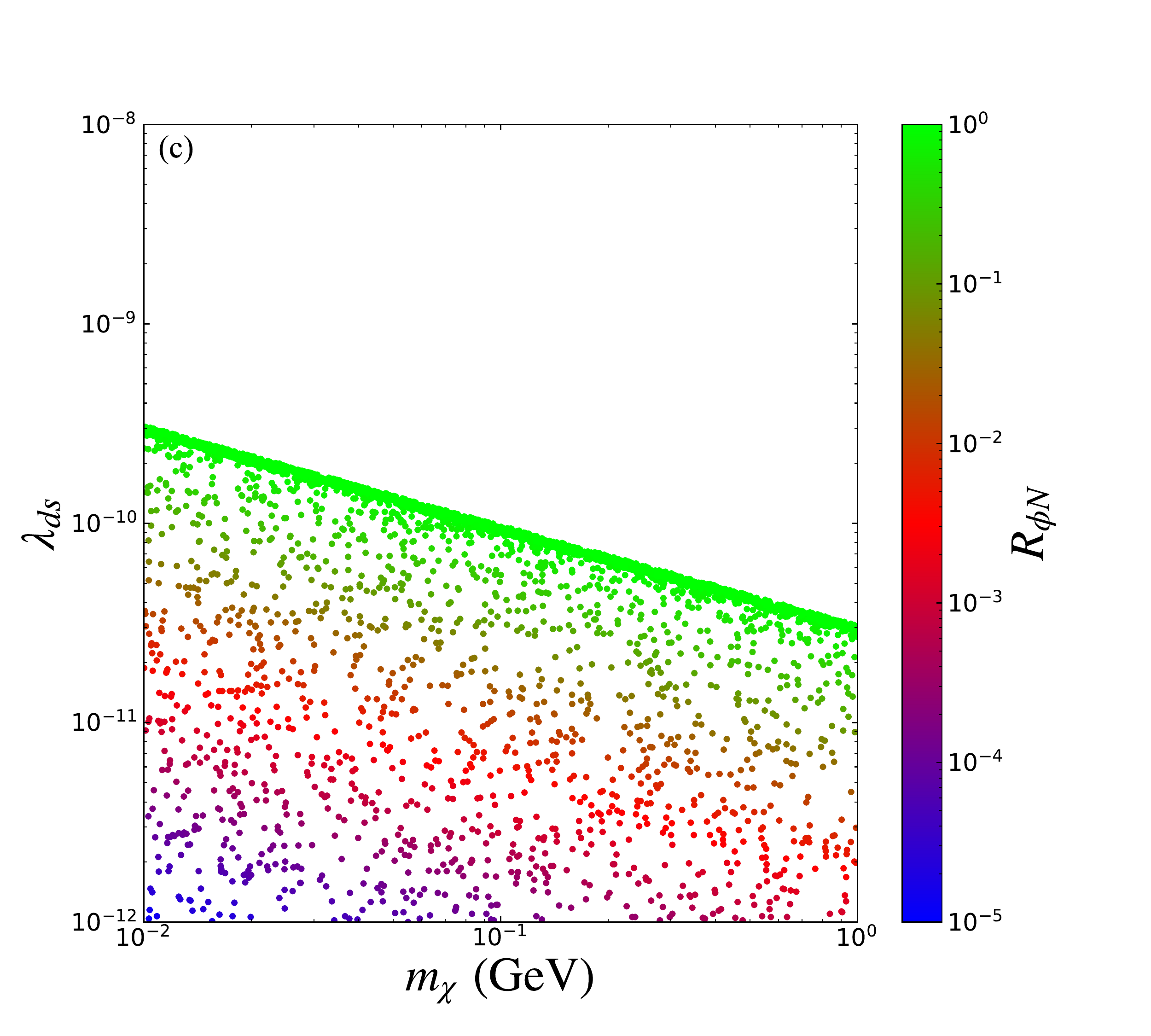}
		\includegraphics[width=0.45\linewidth]{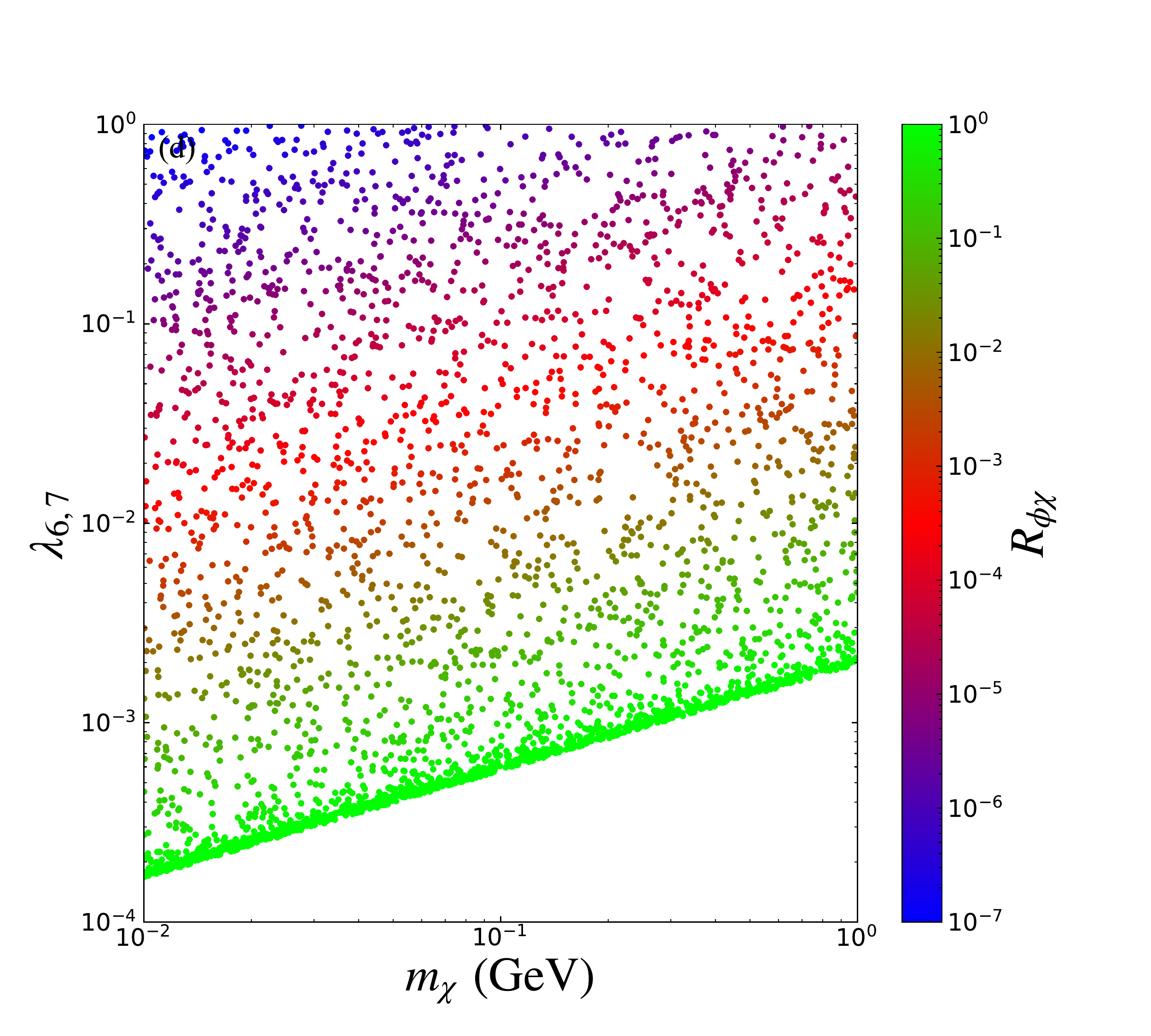}
	\end{center}
	\caption{Allowed parameter space with correct relic density for fermion FIMP DM $\chi$. Panel (a) and (b) correspond to the FIMP scalar $\phi$, and panel (c) and (d) correspond to the WIMP $\phi$. }
	\label{FIG:FRelic2}
\end{figure}

Next, we consider the fermion DM $\chi$. Depending on the quartic coupling $\lambda_{6,7}$, the dark scalar $\phi$ can be either FIMP or WIMP, so a two part scan is also performed here. Assume that $\phi$ is a FIMP, we take the decay mode $N\to \phi\chi$ with $\phi\to \chi\nu$ for illustration. We scan the parameter space
\begin{eqnarray}
	\begin{aligned}
	 m_\chi\in[0.01,1]~\GeV, m_\phi\in[1,100]~\GeV, \lambda_{6,ds}\in[10^{-12}, 10^{-8}], 
	\end{aligned}
\end{eqnarray}
where we have fix $m_N=500$ and $\lambda_6=\lambda_7$. The results are shown in Fig.~\ref{FIG:FRelic2} (a) and (b), where $R_{N\chi}=\Omega_{N\chi}h^2/\Omega_{\rm DM}h^2$ denotes the relative contribution from direct decay $N\to \phi\chi$.  When this channel is the dominant one,  $\lambda_{ds}\in[2.0\times 10^{-11},2.0\times 10^{-10}]$ should be satisfied.  Meanwhile, $R_{\phi\nu}=\Omega_{\phi\nu}h^2/\Omega_{\rm DM}h^2$ represents the relative contribution from direct decay $\phi\to\chi\nu$, where $\phi$ is generated via the quartic scalar couplings $\lambda_{6,7}$. It is worth noting that $\phi$ is mainly produced by the decay of Higgs $h$  when $m_\phi<m_h/2$, so  $\lambda_6=\lambda_7\in[6.5\times 10^{-12},6.5\times 10^{-11}]$ is enough to make this channel dominant. For $m_\phi\geq m_h/2$, $\phi$ is produced via the $2\to2$ scattering of SM particles and $\Phi_\nu$. And $\lambda_6=\lambda_7\in[7.4\times 10^{-11},7.4\times 10^{-10}]$ is required to make the late decay $\phi\to \chi\nu$ dominant.

Then for a WIMP $\phi$, we consider the decay mode $\phi\to N\chi$ and scan the parameter space
\begin{eqnarray}
	\begin{aligned}
	 m_\chi\in[0.01,1]~\GeV, \lambda_6\in[10^{-4}, 1], \lambda_{ds}\in[10^{-12}, 10^{-8}],
	\end{aligned}
\end{eqnarray}
where we have fix $m_\phi=500$ GeV, $m_N=100$ GeV and $\lambda_6=\lambda_7$. The results are shown in Fig.~\ref{FIG:FRelic2} (c) and (d), where $R_{\phi N}=\Omega_{\phi N}h^2/\Omega_{\rm DM}h^2$ represent the relative contribution from dominant freeze-in decay $\phi\to N\chi$ with $\phi$ in the thermal equilibrium state. Contribution of the freeze-in $\phi\to N\chi$ is proportional to $\lambda_{ds}$,  which set an upper bound $\lambda_{ds}\in[3.1\times 10^{-11}, 3.1\times 10^{-10}]$ with $m_\chi$ from 1 GeV to 0.01 GeV. $R_{\phi\chi}=\Omega_{\phi\chi}h^2/\Omega_{\rm DM}h^2$ indicates the relative contribution of the decay of WIMP $\phi$ after freeze-out. $R_{\phi\chi}$ is inversely proportional to $\lambda_{6,7}$. And $\lambda_6=\lambda_7\in[1.6\times 10^{-4}, 2.0\times 10^{-3}]$ with $m_\chi$ from 0.01 GeV to 1 GeV has been obtained by requiring correct relic density.

\subsection{Constraints}

For FIMP DM, its couplings are too small to be detected at the traditional direct and indirect DM detection experiments. We consider a major constraint named free-streaming length for FIMP DM, it represents the average distance a particle travels without a collision \cite{Falkowski:2017uya}
\begin{equation}
	r_{\rm FS}=\int_{a_{\rm rh}}^{a_{\rm eq}}\frac{\langle v\rangle}{a^2H}da
	\approx \frac{a_{\rm nr}}{H_0\sqrt{\Omega_R}}
	\left(0.62+\ln\left(\frac{a_{\rm eq}}{a_{\rm nr}}\right)\right),
\end{equation}
where $a_{\rm eq}$, ${a_{\rm rh}}$ and ${a_{\rm nr}}$  represent Friedmann-Robertson-Walker scale factors in equilibrium, reheating and when DM becomes nonrelativistic, rspectively. We use the approximate expression
\begin{equation}\label{rfs}
	r_{\rm FS} \simeq2.3\times 10^{-6}\left(\frac{\GeV}{m_{\rm DM}}\right)\text{Mpc}
\end{equation}
for a simple estimation.  For DM mass in the range of $[0.01,1]$ GeV, the predicted value is approximately $r_\text{FS}\in[2.3\times10^{-6},2.3\times10^{-4}]$~Mpc. Such a small $r_\text{FS}$ will satisfy the most stringent bound comes from small structure formation $r_\text{FS}<0.1$ Mpc \cite{Boyarsky:2008xj}, therefore all the DM particles in our analysis are cold DM.

In our FIMP DM scenarios, the next lightest stable particle (NLSP) decays into DM and sterile neutrinos $N$ when $N$ is light enough. For $\lambda_{ds}\gtrsim10^{-12}$, NLSP decays fast to avoid constraints from BBN \cite{Cheng:2020gut}. On the other hand, NLSP will decay to DM and light neutrinos through $\chi\to\phi\nu$ or $\phi\to\chi\nu$ when the $N$ is heavier than the dark sector. The tiny mixing angle $\theta$ between sterile and light neutrinos will lead to a delayed decay of NLSP, so the energetic final states neutrinos can be constrained and captured by certain experiments \cite{Bandyopadhyay:2020qpn}. We consider the scenario in Fig.~\ref{FIG:FFeyn} (c) for scalar DM $\phi$ and the scenario in Fig.~\ref{FIG:FFeyn2} (a) for fermion DM $\chi$. Typically, the lifetime of NLSP is about $\tau_{\rm NLSP}\sim 10^{12}$ s with $m_{\rm NLSP}$, $\lambda_{ds}\sim10^{-11}$ and $\theta\sim10^{-7}$, which is close to the time of recombination.

\begin{figure}
	\begin{center}
		\includegraphics[width=0.45\linewidth]{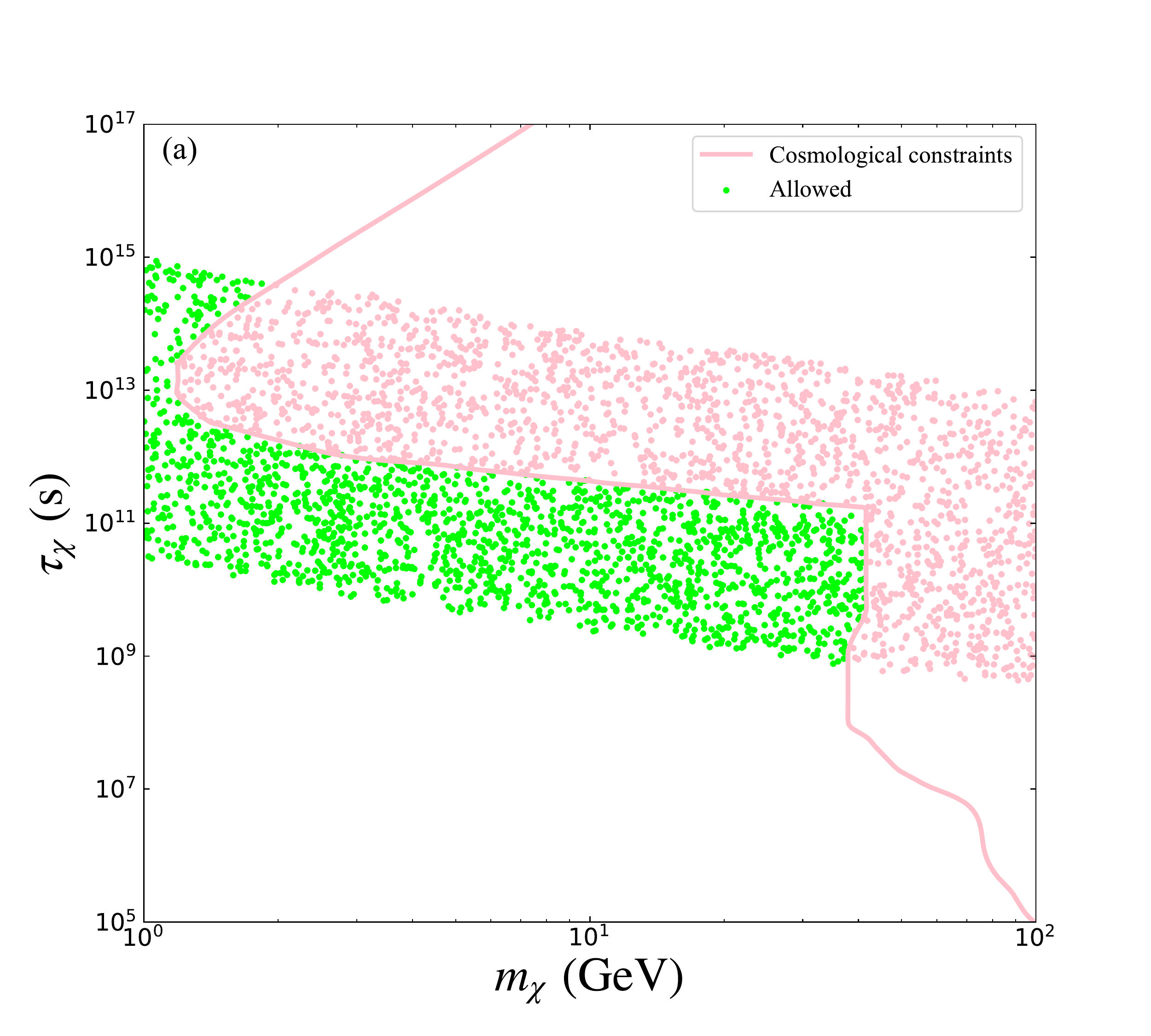}
		\includegraphics[width=0.45\linewidth]{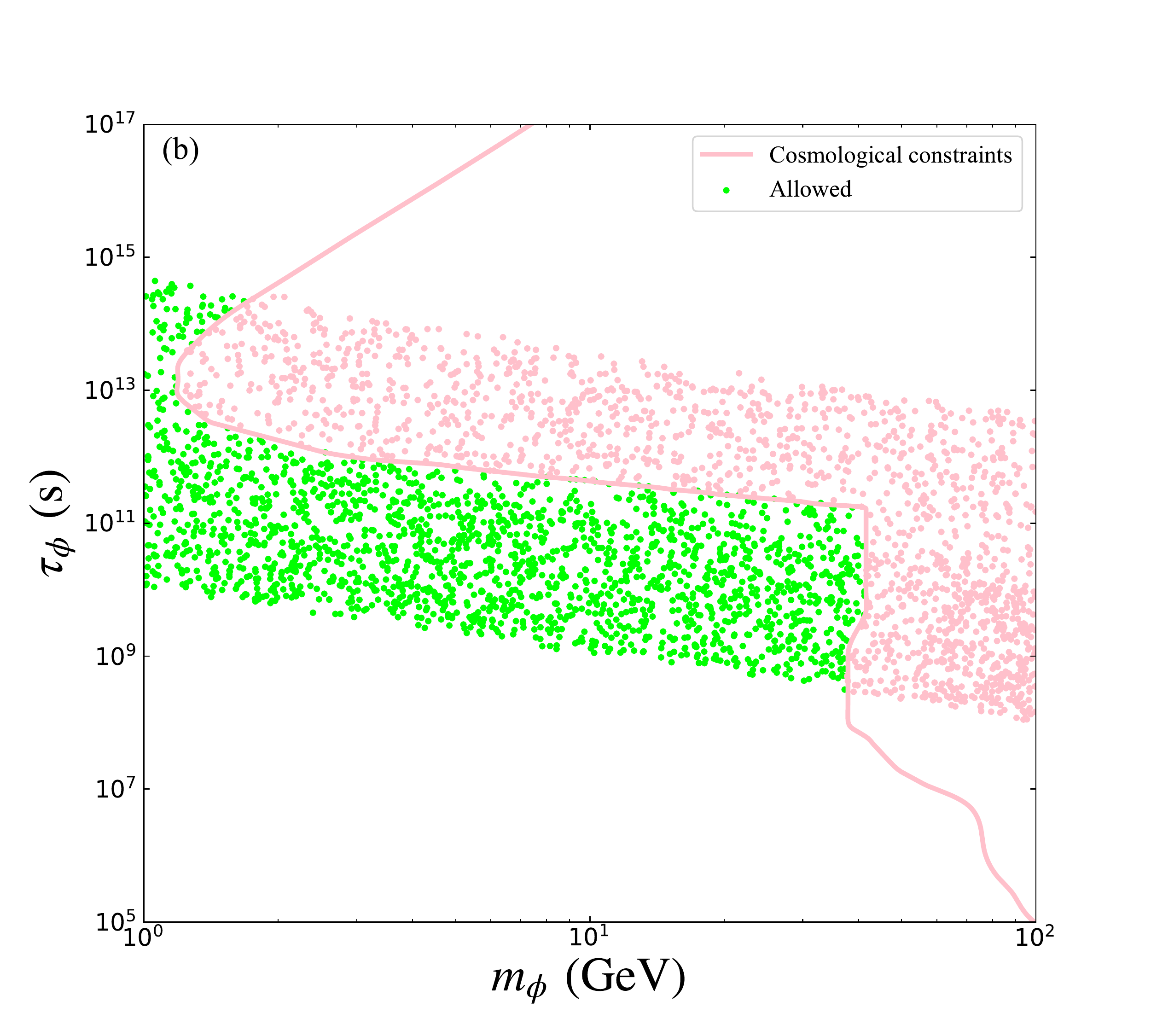}
	\end{center}
	\caption{The cosmological constraints for two specified scenarios. Left panel: DM $\phi$. Right panel: DM $\chi$. The pink curves represent the constraint from CMB and BBN \cite{Hambye:2021moy}. The pink and green points represent the excluded and allowed parameter spaces respectively.}
	\label{FIG:FCc}
\end{figure}

The final state neutrinos from long-lived NLSP can induce electromagnetic and hadronic showers, which will affect the CMB and BBN. In Fig.~\ref{FIG:FCc}, we show the CMB and BBN constraints in the parameter space of $m_{\rm NLSP}$ and $\tau_{\rm NLSP}$ \cite{Hambye:2021moy}, and the scanning results of the considered two scenarios. The cosmological constraints depend on the fractional abundance $f_{\rm NLSP}=\Omega_{\rm NLSP}/\Omega_{\rm DM}$. The constraints in Fig.~\ref{FIG:FCc} correspond to the most stringent one with $f_{\rm NLSP}=10^{4}$, because in the scanning parameter space $f_{\rm NLSP}$ could reach $10^{4}$ in the extreme case when $m_{\rm NLSP}=100$ GeV and $m_{\rm DM}=0.01$ GV.
It is obvious that the mass of NLSP should be less than 40 GeV. And the larger lifetime is usually satisfied only when the mass of NLSP is smaller.

The energetic neutrinos from delayed NLSP decay might be detectable at the neutrino experiments. The neutrino flux from delayed NLSP decay is calculated as \cite{Bandyopadhyay:2020qpn},
\begin{equation}\label{nf}
		\Phi_{\rm cos}\equiv E_\nu^2 \frac{d\varphi}{dE_\nu}=E_\nu\left(\frac{n_\text{NLSP}^0}{\tau_\text{NLSP}}\right)\left(\frac{e^{-t(x)/\tau_\text{NLSP}}}{H(x)}\right)\theta^{'}(x),
\end{equation}
where $E_\nu$ is the observed neutrino energy, $d\varphi/dE_\nu$ is the predicted neutrino flux, $n_{\rm NLSP}^0$ is the NLSP number density when it acts as a decaying particle, $\theta^{'}(x)$ is the Heaviside theta function.   We have $n_{\rm NLSP}^0=\rho_{\rm DM}/m_{\rm DM}$ with the observed DM energy density $\rho_{\rm DM}=0.126\times10^{-5}$ $\rm{GeV/cm^3}$. The cosmic time $t(x)$ at red-shift $1+x$ and the Hubble parameter $H(x)$ in the standard cosmology are given by
\begin{eqnarray}\label{nf2}
t(x)&\approx& \frac{4}{3H_0}\left(\frac{\Omega_{\rm r}^{3/2}}{\Omega_{\rm m}^{2}}\right)\left(1-\left(1-\frac{\Omega_{\rm m}}{2(1+x)\Omega_{\rm r}}\right)\sqrt{1+\frac{\Omega_{\rm m}}{(1+x)\Omega_{\rm r}}}\right),  \\
H(x)&=&H_0\sqrt{\Omega_\Lambda+(1+x)^3\Omega_{\rm m}+(1+x)^4\Omega_{\rm r}},
\end{eqnarray}
where $x=E_0/E_\nu-1$ with initial energy of NLSP $E_0=(m_{\rm NLSP}^2-m_{\rm DM}^2)/2m_{\rm NLSP}$, the Hubble constant $H_0=100h$ $\rm{km/s/Mpc}$ with $h=0.6727$ \cite{Planck:2018vyg}, the dark energy, matter and radiation (CMB photons and neutrinos) fractions are $\Omega_\Lambda=0.6846, \Omega_{\rm m}=0.315$ and $\Omega_{\rm r}=9.265\times10^{-5}$.

\begin{figure}
	\begin{center}
		\includegraphics[width=0.6\linewidth]{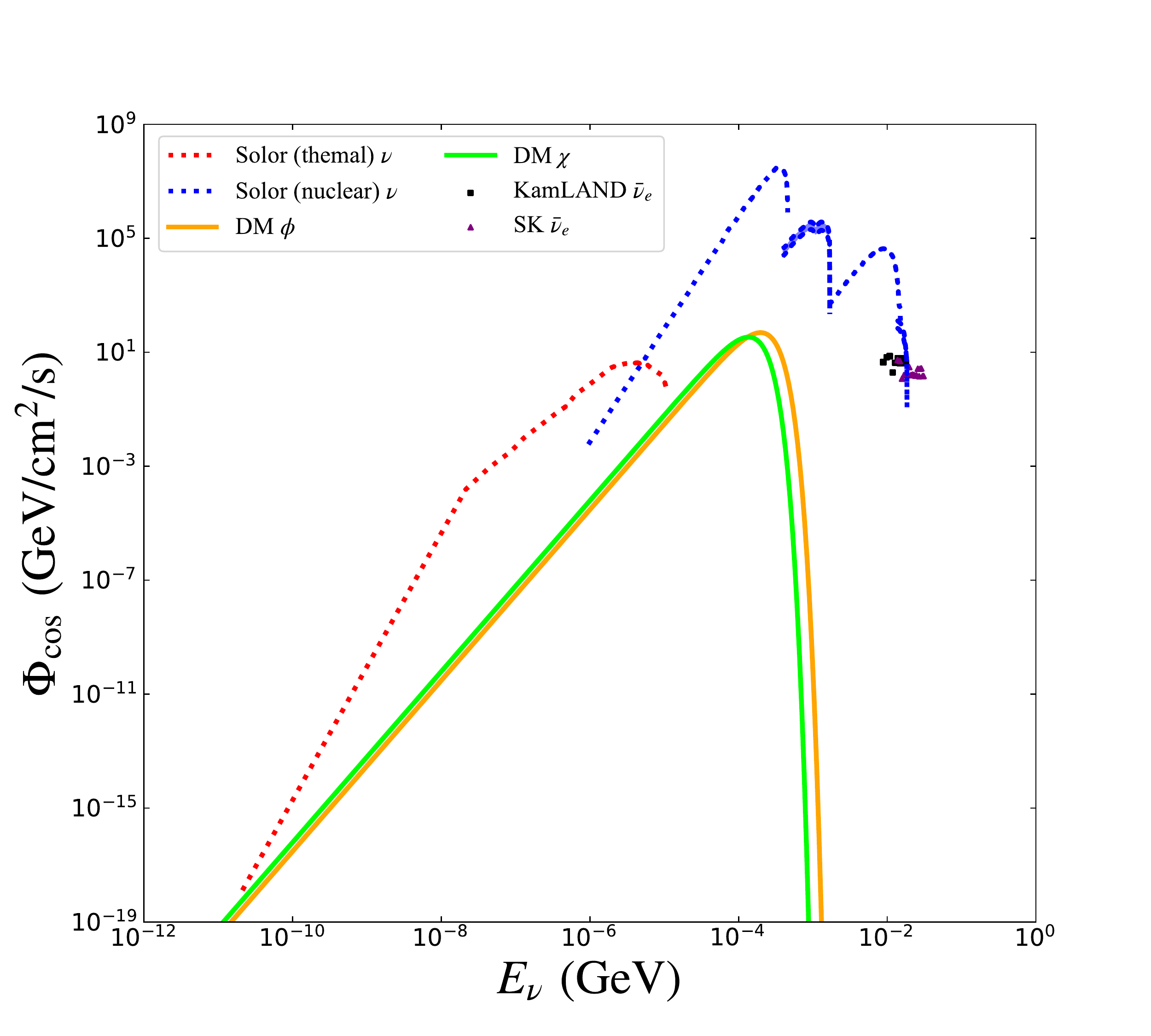}
	\end{center}
	\caption{The results of neutrino flux constraints for two benchmark points. The cyan and green solid lines represent the results from DM $\phi$ and $\chi$, respectively. The red dotted line represent the thermal solar neutrinos flux \cite{Vitagliano:2019yzm}. The blue dotted lines are the solar neutrino flux for nuclear production processes \cite{Vitagliano:2019yzm}. The orange squares and purple triangles represent diffuse supernova neutrino background (DSNB) flux of the electron anti-neutrinos with the KamLAND \cite{KamLAND:2011bnd} and SK \cite{Super-Kamiokande:2013ufi} data, respectively. }
	\label{FIG:FNf}
\end{figure}

The predicted neutrino flux is shown in Fig. \ref{FIG:FNf}. For both scalar and fermion DM, we assume $m_{\rm NLSP}=10$ GeV, $m_{\rm DM}=0.1$ GeV. These two benchmark points meet $\tau_\chi=2.38\times10^{10}$ s for DM $\phi$ and $\tau_\phi=1.16\times10^{10}$ s for DM $\chi$, which are allowed by the cosmological constraints in Fig.~\ref{FIG:FCc}. Although the initial neutrino energy from NLSP decay is $E_0\sim m_{\rm NLSP}/2=5$ GeV, the observed energy $E_\nu$ is red-shifted to below 1 MeV, which makes these neutrinos are hard to detect at current neutrino experiments. 

\section{Conclusion} \label{SEC:CL}  

In this paper, we study the phenomenology of sterile neutrino portal DM in the $\nu$THDM. This model introduces one neutrinophilic scalar doublet $\Phi_{\nu}$ and sterile neutrinos $N$. The soft-term $\mu^2 \Phi^\dag \Phi_\nu$ induces a tiny VEV for $\Phi_\nu$, which leads to naturally tiny neutrino mass with $\Phi_\nu$ and $N$ around TeV scale. In this way, the Yukawa coupling between $\Phi_{\nu}$ and $N$ is large enough to keep $N$ in thermal equilibrium. The dark sector is consist of one Dirac fermion singlet $\chi$ and one scalar singlet $\phi$, which are odd under a $Z_2$ symmetry. The sterile neutrinos $N$ are the mediator between the DM and SM. Depending on the coupling strength, the DM can be either WIMP or FIMP. 

For the WIMP scenario, the scalar candidate $\phi$ can annihilate into SM final states, into neutrinophilic scalars, and into sterile neutrinos. The Higgs portal interaction is tightly constrained by direct detection.  Under the constraints from Higgs invisible decay, direct and indirect detection, $m_\phi$ should be larger than  about 60~GeV. For the fermion candidate $\chi$, $\chi\bar{\chi}\to NN$ is the only possible annihilation channel. Current Higgs invisible decay and direct detection limits do not exclude any samples. Meanwhile, the indirect detection has exclude $m_\chi\lesssim 60$ GeV. In the future, the indirect detection experiment as CTA is able to probe the region with DM mass less than 1 TeV for both scalar and fermion candidate.

For the FIMP scenario, we consider the direct production of DM from freeze-in mechanism, as well as contributions from late decay of NLSP. For scalar FIMP, it can be generated from decay of SM Higgs boson $h$ or sterile neutrinos $N$, annihilation of neutrinophilic scalars via $\Phi_\nu\Phi_{\nu}\to \phi\phi$  or via $\Phi_{\nu} L\to \phi \chi$. For fermion FIMP, it is mainly generated from decay of sterile neutrino $N$ or NLSP $\phi$. Both the FIMP and WIMP case of $\phi$ are discussed. Although the traditional direct and indirect DM detection experiments are hard to probe FIMP DM, the energetic neutrinos from delayed decay of NLSP will lead to constraints from CMB, BBN, and neutrino experiments when the sterile neutrinos are heavier than the dark sector. Under these constraints, the NLSP mass should be less than about 40 GeV.

\section*{Acknowledgments}
This work is supported by the National Natural Science Foundation of China under Grant  No. 11975011, 11805081 and  11635009, Natural Science Foundation of Shandong Province under Grant No. ZR2019QA021 and ZR2018MA047.


\end{document}